\documentclass{aastex61}
\usepackage{amsmath}
\usepackage{natbib}
\begin{document}

\title{Analytic Formulation of 21~cm Signal from Cosmic Dawn: Ly$\alpha$ fluctuations}

\author{Janakee Raste}
\affiliation{Raman Research Institute, Bangalore 560080, India} 
\affiliation{Joint Astronomy Program, Indian Institute of Science, Bangalore 560012, India}

\author{Shiv Sethi}
\affiliation{Raman Research Institute, 
Bangalore 560080, India}

\begin{abstract}

We present an analytic formalism to compute the fluctuating component of the \ion{H}{1} signal and extend it  to take into account the effects of partial Ly$\alpha$ coupling during the era of cosmic dawn.  We use excursion set formalism to calculate the size distribution of randomly distributed self-ionized regions. These ionization bubbles are surrounded by  partially heated and Ly$\alpha$  coupled regions, which create spin temperature $T_S$ fluctuations. We use the ratio of number of Ly$\alpha$ to ionizing photon ($f_L$) and number of X-ray photons emitted per stellar baryons ($N_{\rm heat}$) as modeling parameters. Using our formalism, we compute the global \ion{H}{1} signal, its autocorrelation and power spectrum in the redshift range $10 \le z \le 30$ for the $\Lambda$CDM model. We check the validity of this formalism for various limits and simplified cases. Our results agree reasonably well with existing results from $N$-body simulations, in spite of following a  different approach and requiring orders of magnitude less computation power and time. We further apply our formalism to study the fluctuating component corresponding to the recent EDGES observation that shows an  unexpectedly deep absorption trough in global \ion{H}{1} signal in the redshift range $15 <z< 19$. We show that, generically, the EDGES observation predicts larger signal in this redshift range but smaller signal at higher redshifts. We also explore the possibility of  negative real-space autocorrelation  of spin temperature and show it can be achieved  for partial Ly$\alpha$ coupling in many cases corresponding to simplified models and complete model without density perturbations. 

\end{abstract}


\section{Introduction} \label{sec:intro}

Our current understanding of the history of the universe suggest that the dark ages of the universe ended around redshift $z \sim 35$ with the formation of first large-scale structures (epoch of cosmic dawn). These collapsed structures emitted radiation  which  heated and ionized their surrounding medium (epoch of reionization -- EoR) by $z \sim 8$ (\cite{Barkana:2000fd,21cm_21cen,2014PTEP.2014fB112N,2010ARA&A..48..127M}). The physics of first stars and galaxies is only partially understood theoretically and poorly constrained with observations. The current observational bound on the cumulative history of reionization is provided by the detection of Gunn-Peterson effect, which indicate  that the universe was fully ionized by $z \simeq 6$. The CMB temperature and polarization anisotropy detections by WMAP and Planck gives the redshift of reionization,  $z_{\rm reion} = 7.75 \pm 0.73$ (\cite{Planck2013,Hinshaw:2012aka,Planck2015,Fan:2000gq,2001AJ....122.2850B,Planck2018}).

The cleanest   probe of  the physics of EoR is through the detection of redshifted hyperfine 21~cm line of neutral hydrogen (\ion{H}{1}). This signal carries crucial information about the first sources of radiation in the universe and their spectrum in three frequency bands: ultraviolet (UV) radiation (that ionizes the surrounding medium), Ly$\alpha$ radiation (that determines the relative population of   neutral hydrogen atoms in hyperfine states), and X-ray photons (which heat and partially ionize the medium). In addition, the sources that  emitted soft radio photons  would also affect the observable \ion{H}{1} signal considerably (e.g. \cite{2018ApJ...868...63E,feng18}). Along with primordial density perturbations given by the $\Lambda$CDM model, the inhomogeneities of these radiation fields establish the length scales of the fluctuating component of the signal. 

The epoch of cosmic dawn and  EoR has been  studied in detail  using numerical, semi-analytic   and,  analytic methods (e.g. \cite{2007MNRAS.376.1680P,2013MNRAS.435.3001T,2013MNRAS.431..621M,LateHeat2,2015MNRAS.447.1806G,2014MNRAS.443..678P,2017MNRAS.464.3498F,2012Natur.487...70V,21CMFAST}). Theoretical estimates based on standard thermal and ionization history  suggest  the global signal is observable in both absorption and emission with its strength in the range $-200\hbox{--}20 \, \rm mK$ in a frequency  $50\hbox{--}150 \, \rm MHz$,  corresponding  to a redshift range $25 > z > 8$ (e.g. \cite{1997ApJ...475..429M,2000NuPhS..80C0509T,2004ApJ...608..611G,Sethi05}). The fluctuating component of the signal is expected to be an order of magnitude smaller on scales in the range $3\hbox{--}100 \, \rm Mpc$, which implies  angular scales  $\simeq 1\hbox{--}30$~arc-minutes (e.g. \cite{ZFH04,FZH04a,FZH04b,2007MNRAS.376.1680P}; for comprehensive reviews see e.g. \cite{21cm_21cen,2014PTEP.2014fB112N,2010ARA&A..48..127M}). Many of the ongoing and upcoming experiments have  the capability to detect this signal in hundreds of hours of integration (e.g. \cite{2015aska.confE...3A,2014MNRAS.439.3262M}). Upper limits on the fluctuating component of the \ion{H}{1}  signal have  been obtained by many ongoing experiments---GMRT, MWA, PAPER, and LOFAR (\cite{2017ApJ...838...65P,2016ApJ...833..102B,2015ApJ...809...61A,2013MNRAS.433..639P}), with the best upper limits of $\simeq (50 mK)^2$ for $k\simeq 0.1 \, \rm Mpc^{-1}$. 

The recent detection of a broad global absorption trough of strength $500 \, \rm mK$ by the  EDGES group (\cite{EDGES2018}) at $\nu \simeq 80 \pm 10 \, \rm MHz$ is the only positive detection of \ion{H}{1} signal at high redshifts.  Observational projects that  are attempting  to detect the global signal (SARAS (\cite{2018Singh1, 2018Singh2}), LEDA (\cite{2015ApJ...799...90B,2018MNRAS.478.4193P}), BIGHORNS (\cite{2015PASA...32....4S}) and SCI-HI (\cite{2014ApJ...782L...9V})) and its fluctuating component (HERA, LOFAR, MWA) might  provide more insight into  the physics of EoR in the  near future. If confirmed, the unexpectedly deep absorption trough detected by EDGES will also  open avenues to investigate  exotic physics (e.g. \cite{Barkana2018,2018arXiv180405318L,2018PhLB..785..159F,2018Natur.557..684M}).

Numerical simulations can provide us insight into the morphology and evolution of the sources in the early universe. However, given the uncertainty in the astrophysics of this epoch, it is useful to develop fast analytic methods which can analyze the signal for  a  large range of scales for  different combinations of  physics inputs and modelling parameters. 
In our previous work (\cite{RS18}, RS18 from now on), we developed a formalism to  analytically compute the autocorrelation and power spectrum of \ion{H}{1} signal in the early phase of cosmic dawn and EoR, when the medium is partially heated and ionized. For simplicity, we had assumed a complete coupling of hydrogen spin temperature $T_S$ with matter kinetic temperature $T_K$. In this paper, we expand the formalism to include the effect of inhomogeneous Ly$\alpha$ coupling on the \ion{H}{1} signal. We also apply our method to study the fluctuating signal which would correspond to the global \ion{H}{1} signal observed by EDGES group. 

In the next section, we review the \ion{H}{1} signal from the EoR and the expected  photoionization and heating of the IGM due to UV and X-ray. In section~\ref{sec:lyalpha}, we model  fluctuations in the signal due to inhomogeneous Ly$\alpha$ coupling. In section~\ref{autoh1sig}, we briefly present the formalism for computing the two-point correlation function of the \ion{H}{1} signal and discuss a few approximations. In section~\ref{sec:res}, we present our results and explore their dependence on our modelling parameters. We summarize the derivation of our formalism in the Appendix~\ref{sec:comp}. We present our conclusions in Section~\ref{sec:sumcon}. 
Throughout this paper, we assume the spatially-flat $\Lambda$CDM model with the following parameters: $\Omega_m = 0.310$, $\Omega_B = 0.049$, $h = 0.677$ and $n_s = 0.967$, with the overall normalization corresponding to  $\sigma_8 = 0.808$ (\cite{Planck2018}). 

\section{Cosmic Dawn and Epoch of Reionization}\label{sec:h1sig} 
The hyperfine splitting of  neutral hydrogen in its ground state causes an energy difference corresponding to   wavelength $\lambda = 21.1 \, \rm cm$. The spin (or excitation) temperature of this line, $T_S$, is a function of ratio of atoms in two hyperfine states. This ratio is determined  by three processes in the  early universe: absorption and stimulated emission  of  CMB photons at a temperature $T_{\rm CMB}$, collisions with atoms and charged particles, and the mixing of hyperfine  levels owing to Ly$\alpha$ photons (Wouthuysen-Field effect).  $T_S$ can be expressed in terms of the colour temperature of Ly$\alpha$ photons, $T_{\alpha}$, gas kinetic temperature $T_K$, and $T_{\rm CMB}$ (\cite{Field1958,21cm_21cen}) as:
	\begin{equation}
		T_S=\frac{T_{\rm CMB}+y_{\alpha}T_{\alpha}+y_c T_K}{1+y_{\alpha}+y_c}
                \label{eq:tsbas}
	\end{equation}
Here $y_c \propto n_{\rm H}, n_e$ (number density of neutral hydrogen atom or electrons; neutral atoms dominate for small ionized fraction) and $y_\alpha \propto n_\alpha$ (number density of Ly$\alpha$ photons). During the dark ages, $1000 < z < 100$, $T_S$ relaxes to $T_{\rm CMB}$. After matter thermally decouples from CMB, $100 < z < 30$, collisions couple $T_S$ to matter temperature $T_K$. When the first sources of radiation form during cosmic dawn, the production of Ly$\alpha$ photons couple the spin temperature to the colour temperature of Ly$\alpha$ $T_\alpha$, which is relaxed to $T_K$ due to multiple scattering of Ly$\alpha$ photons with \ion{H}{1} (e.g. \cite{2004ApJ...602....1C,1959ApJ...129..551F,1994ApJ...427..603R}). If, at any redshift, $y_{\text{tot}}=y_c+y_{\alpha} \gtrsim T_{\rm CMB}/T_K $, then $T_S$ is strongly coupled to $T_K$. Otherwise, in absence of these coupling mechanisms, it relaxes to $T_{\rm CMB}$.

The \ion{H}{1} is observable in emission or absorption depending on whether its spin temperature $T_S$ is greater than or less than $T_{\text{CMB}}$. The CMB spectral distortion caused by this effect is observable and can be expressed as (e.g. \cite{21cm_21cen,1997ApJ...475..429M,1999A&A...345..380S,2004ApJ...608..611G,Sethi05}):
	\begin{align}
		\Delta T_b & \simeq 26.25\;n(1+\delta)\left(1-\frac{T_{\text{CMB}}}{T_S}\right) \left(\frac{1+z}{10}\frac{0.14}{\Omega_m h^2}\right)^{\frac{1}{2}} \left(\frac{\Omega_b h^2}{0.022}\right) \text{mK} \label{overallnorm}
	\end{align}
Here the redshift space distortion is ignored. We have expressed \ion{H}{1} number density as, $n_H = \bar n_H n (1+\delta)$ and the mean density $\bar n_H$ has been absorbed in the prefactor of Eq.~(\ref{overallnorm}). $\delta$ is overdensity of the gas. We have assumed that a small volume at any point is either completely neutral or completely ionized, therefore  a variable $n$  is defined which is unity if the medium is neutral and zero otherwise. We further  define dimensionless temperature fluctuation as (\cite{ZFH04}):
	\begin{align}
		\psi &= n(1+\delta)(1-s), \label{psidef}
	\end{align}
which captures the density ($\delta$), ionization ($n$) and spin temperature, $T_S$,  inhomogeneities. Here  $s = T_{\text{CMB}}/T_S$. These quantities are functions of position, and thus they contribute to the spatial fluctuation of the signal. We suppress this dependence for notational clarity.

At the end of the dark ages, high density regions of the universe collapse and form structures of a range of masses. The radiation emitted by them change the properties of their surrounding medium. In our work we assume, that the smallest mass that  can collapse corresponds to  \ion{H}{1}-cooled halo (e.g. \cite{Barkana:2000fd,Dayal2018}):
	\begin{align}
		M_\text{min} = 3.915 \times 10^8 \frac{1}{\Omega_m^{1/2}h\;(1+z)^{3/2}} M_\odot
	\end{align}
We consider fluctuations of hydrogen-ionizing, Ly$\alpha$, and X-ray radiation fields emitted by the sources on the brightness temperature inhomogeneities.  

\subsection{Photoionization} \label{sec:photion}
The hydrogen ionizing (ultraviolet -- UV) photons emitted from the star within the collapsed structures are absorbed in the immediate vicinity of the sources and carves out \ion{H}{2} regions around them in the Intergalactic Medium (IGM). We use excursion set formalism to compute size distribution of the ionization bubbles by defining self-ionized regions (\cite{FZH04a}). Such region have enough sources to ionize all gas within them. Their sizes are determined by the ionization efficiency factor, defined as $\zeta = 1/f_{\text{coll}}=M_\text{tot}/M_\text{coll}$. Here, $f_{\rm coll}$ is the fraction of collapsed mass inside the self-ionized region. $\zeta$ is a function of property of the sources as well as the surrounding halo (\cite{FZH04a}),
	\begin{equation}
        \zeta = f_\star f_{\text{esc}}N_{\text{ion}}N_{\text{rec}}^{-1}.
	\label{eq:defzeta}
    \end{equation}
Here, $f_\star$ is the fraction of collapsed baryons that is converted into stars and $f_{\text{esc}} $ is the fraction of ionizing photons that escape the source halo. $N_{\text{ion}}$ is number of UV photons created per stellar baryon, while  $N_{\text{rec}}$ is the number of recombinations. 
We assume $\zeta$ to be independent of redshift in this paper, even though it can evolve with time owing to the evolution of quantities used to define it. For higher value of $\zeta$, the reionization is completed at higher redshift (e.g. RS18). These self-ionized regions are larger than the \ion{H}{2} regions of a single source, since they are created by highly clustered multiple sources in the early universe (for $\Lambda$CDM model). For our work, we assume the region to be spherical. 

\subsection{X-Ray Heating} \label{sec:xrayheat}

Photons of energy $E \gg 13.6 \, \rm eV$ (X-rays) escape the \ion{H}{2} region into the surrounding medium. These photons ionize the neutral gas upto 10\% and can impart upto 20\% of their energy into heating the gas through photoionization and secondary collisional ionization and excitation (e.g. \cite{1985ApJ...298..268S,Heating2001}). In our study, we have assume the medium outside the ionized region to be completely neutral ($n=1$) since the fraction of ionization due to this process is generally small.

The photoionization cross section of X-ray photons falls as $\sigma_i(\nu) = {\sigma_i}_0 (\nu/\nu_i)^{-3}$, with $\nu_i$ being the ionization threshold frequency of species $i$. In this work, we only consider two species: neutral hydrogen and neutral helium with their relative fractions $x_i=12/13$ and $1/13$ of the baryon number, respectively. This is a valid assumption since the metallicity in early universe is very low. As the low energy X-ray photons are absorbed with higher probability, they contribute to heating the medium immediately surrounding the \ion{H}{2} region, whereas the higher energy photons free-stream through the medium and might get absorbed far away from any source. These photons uniformly heat up the whole IGM to some background temperature $T_{\text{bg}}$.

We assume the X-ray photon source luminosity to be given by a power law (e.g. \cite{21CMFAST} and references therein), $\dot{N}_\nu = \dot{N}_t (\nu/\nu_\text{min})^{-\alpha}$, where $\nu_\text{min}$ is the lowest frequency of X-ray photons escaping from source halos. $N_{\text{heat}}$  is the number of X-ray photons emitted per stellar baryons. We assume that $f_H =0.15$ is  the fraction of energy of emitted photoelectron that goes into heating the medium (\cite{1985ApJ...298..268S}, \cite{Heating2001}). Other than adiabatic expansion of the universe, we neglect all other cooling processes. 

We divide the neutral hydrogen volume in two zones. In the {\it near zone} the heating is dominated by X-ray photons from an individual self-ionized region. In the {\it far zone}, the contribution from all the far away background sources is taken into account. For more details and explanations, see RS18, where the increase in temperature due to a self-ionized region of radius $R_x$ at a distance $R_0$ from the centre of the ionized region was calculated at redshift $z_c'$,
	\begin{align}
		\Delta T' &= \frac{h f_H \alpha N_{\text{heat}} f_\star n_0 \nu_\text{min}^{\alpha}}{3 k_B\zeta} \frac{ R_x^3}{R_0^2} (1+z'_c)^2 \nonumber \\
				& \quad \quad \int_{t(z_\star)}^{t(z_0)} \mathrm{d}t'\frac{f_i(t')}{f_i(t)} \frac{\dot{f}_{\text{coll,g}}}{f_{\text{coll,g}}} \left(\frac{1+z'}{1+z}\right)^{\alpha+1} \nonumber \\
				& \quad \quad \quad \int_{\nu_\text{min}'}^{\infty} \mathrm{d}\nu' {\nu'}^{-\alpha-4} \mathrm{e}^{-\tau(R_0,\nu')} \sum_i (\nu'-\nu_i)  x_i {\sigma_i}_0{\nu_i}^3 
		\label{eq:fintemp}
	\end{align}
The primed quantities are calculated at the receiving point (point $P$), unprimed quantities are at the source (point $S$), and quantities with $0$ subscript are comoving quantities. $\alpha$ is the X-ray spectra power law index and $f_i$ are global ionization fraction.
	
\subsection{Ly$\alpha$ radiation} \label{sec:lyalpha}
For EoR studies, all the radiation between Ly$\alpha$ and Lyman-limit is referred to as Ly$\alpha$ radiation emitted from the source and we shall follow this convention. The Ly$\alpha$ contribution at any point arises from  two main factors: Ly$\alpha$ emitted from the sources and Ly$\alpha$ created due to X-ray photo-electrons (\cite{Heating2001}). The latter is generally small and we neglect it in this paper.

Ly$\alpha$ photons emitted from the sources escape the surrounding \ion{H}{2} region and redshift until their frequency nearly equals the resonant frequency of one of the Lyman series lines. When a photon redshifts to the frequency corresponding to one of the Lyman series lines, it gets scattered by the neutral hydrogen \footnote{The scattering cross-section falls with increasing $n$ so this scenario is applicable if the optical depth of scattering  in the expanding medium exceeds unity. The requirement is readily met for the transitions of interest  ($n <20$)  in the paper. } and eventually cascades to frequency corresponding to Ly$\alpha$. Given the complicated frequency structure of Lyman series lines, these photons are absorbed at varying distances from the source. Thus, the coupling depends on two factors: the region of influence of the Ly$\alpha$ radiation and the coupling coefficient $y_\alpha$.
Photons between Ly$\alpha$ and Ly$\beta$ frequencies can be absorbed by $H_2$ molecule, but we ignore this effect as density of $H_2$ is very low in IGM. 

We define Ly$n$ influence region as the distance traveled by the Ly$(n+1)$ photons to redshift to  Ly$n$ frequency. If these photons were emitted at $z=z_e$ and absorbed at $z = z_a$ with $\nu_e = \nu_{n+1}$ and and $\nu_a = \nu_n$, then, the comoving distance traveled by the photon before it is absorbed in an expanding universe is ($n \ge 2$): 
	\begin{align}
		R_{\rm max} (n) \simeq \frac{16040 \;\;\text{Mpc}}{(1+z_e)^{1/2}} \left[ \left( \frac{n^3(n+2)}{(n+1)^3(n-1)} \right)^{1/2} -1 \right] 
\label{eq:lyainf}
	\end{align}
These influence regions become smaller with increasing $n$. We note that  $R_{\rm max}(2)$ (Ly$\alpha$ influence region) is much larger than the mean distance between ionization bubbles at any redshift. For $\zeta=7.5$, the values of mean comoving distance between bubbles for redshift 25, 20, 15 is 7.85 Mpc, 2.29 Mpc and 0.96 Mpc respectively. 
Therefore, Ly$\alpha$ regions are very large and merge very early. However, this  would create homogeneous coupling to \ion{H}{1} atoms only if Ly$\alpha$ coupling coefficient, $y_\alpha$ is high enough (Eq.~(\ref{eq:tsbas})). $y_\alpha$ is a function of Ly$\alpha$ photon (physical) number density $n'_{\alpha}$ (\cite{Field1958,2004ApJ...602....1C}):
	\begin{align}
		y_\alpha &= \frac{ \sqrt{2}h c^4}{36 \pi \nu_\alpha^2 k} \sqrt{\frac{m_H}{2kT_K}} \frac{A_\alpha}{A_{10}} \frac{\nu_{10}}{\nu_\alpha} \frac{n'_{\alpha}}{T_K} \label{eq:lymancoup} 
	\end{align}
To calculate the number density of Ly$\alpha$ photons ($n'_{\alpha}$) received at any point from ionizing sources, we use the method used to calculate X-ray heating in RS18. Assuming  flat spectrum between Ly$\alpha$ and Lyman-limit, the number density of Ly$\alpha$ photons at a comoving distance $R_0$ from the source is,
	\begin{align}
		n'_{\alpha,\star} = \frac{\dot{N}_\alpha}{4\pi c R_0^2} \frac{2\Delta \nu_\alpha}{\nu_\beta-\nu_\alpha}\frac{(1+z')^3}{1+z} . \nonumber
	\end{align}
Here $\Delta \nu_\alpha = \sqrt{2kT_K/m_H c^2} \nu_\alpha$ is the  Doppler line width. This factor arises because at the source the photons are emitted with frequencies between $\nu_\beta$ and $\nu_\alpha$, but the only frequencies which are absorbed at redshift $z'$ are in the range of $\Delta \nu_\alpha$ around $\nu_\alpha$.  
The Ly$\alpha$ luminosity,  $\dot{N}_\alpha$, can be expressed in terms of the  size of ionization halo, assuming that the Ly$\alpha$ luminosity scales with ionizing luminosity with a factor $f_L$. Using the balance between ionization and recombination in the ionizing region,
	\begin{align}
		\dot{N}_\alpha = f_L\frac{4\pi}{3} n_0^2 \alpha_B C  R_x^3 (1+z)^3. \nonumber
	\end{align}
It would be reasonable to assume that, $f_L >1$, because Ly$\alpha$ photons escape the halo more easily than ionizing photons. However, for the sake of completeness, we take $0.1 < f_L < 1000$ in this paper. Combining all these, we get,
	\begin{align}
		y_{\alpha,\star} T_K &\simeq \frac{S_\alpha}{54 \pi}\frac{h c^2 A_\alpha \nu_{10} n_0^2 \alpha_B}{\nu_\alpha^2 k A_{10}(\nu_\beta-\nu_\alpha)} f_L  C \frac{R_x^3}{R_0^2}(1+z')^3(1+z)^2
		\label{eq:lymanstar}
	\end{align}
Here, $S_\alpha$ is a correction factor of order unity as defined in \cite{2004ApJ...602....1C}, which depends on the photon spectrum around Ly$\alpha$ line (\cite{2006MNRAS.367..259H}, \cite{21cm_21cen}). This expression has been derived in somewhat different manner in \cite{21cm_21cen}.	

We find contributions to $y_\alpha$ from higher order Lyman transitions (eg. from Ly$\gamma$ to Ly$\beta$) in a similar way, by counting the number of Ly$n$ influence regions the point falls within. The effect of higher order transitions is expected to be subdominant since, for a continuum source,  the total number of photons between Ly$\beta$ to Lyman-limit is smaller than in the frequency range  Ly$\beta$ and Ly$\alpha$. However, these photons can have substantial impact near an ionizing source since they are absorbed closer to the source given their smaller influence regions. 
If the distance of a point from the source $R_0$ is such that, $R_{\rm max}(n+1)<R_0<R_{\rm max}(n)$, then the point in question will have $n$ Ly$\alpha$ photons in its vicinity rather than one photon if only transition between Ly$\alpha$ and Ly$\beta$ is considered.
This means that Ly$\alpha$ flux from the source centre generally falls more rapidly than $1/r^2$ when this effect is taken into account.  In this paper, we only consider Lyman series lines which have influence regions larger than the ionization bubble radius \footnote{Photons with influence regions smaller than the ionization bubble will redshift to lower and lower Lyman series lines until they cross the ionization region boundary. However, their effect would be very small and very close to the boundary of the ionization region. It is not useful to model them in more detail, since, several other assumptions would break down so close to the boundary (e.g. sphericity of bubbles, sharp boundary of ionization regions).}.

\subsection{Collisional Coupling}
The collisional coupling of spin temperature $T_S$ and matter kinetic temperature $T_K$ due to scattering of neutral  hydrogen and electrons (Eq~\ref{eq:tsbas}) can play an important role at lower redshift ($z \leq 30$) too. The coupling coefficient is proportional to the number density of colliding particles, 
	\begin{align}
		y_c = \frac{(n_H k_{10}^H+n_e k_{10}^e)}{A_{10}}\frac{T_\star}{T_K}. \label{eq:ycdef}
	\end{align}
For collision rate coefficients, we use the following fits (\cite{Zygelman}, \cite{21cm_21cen}): 
	\begin{equation}
		k_{10}^e=\left\{
		\begin{array}{cl}
			{\rm exp}\left(-9.607+0.5 {\rm log}(T_K) {\rm exp}(-\frac{({\rm log}(T_K))^{4.5}}{1800})\right)\;\;{\rm cm}^3{\rm s}^{-1} &\quad T_K \leq 10^4 {\rm K}\\
			k_{10,e}(T_K = 10^4 {\rm K}) &\quad T_K > 10^4 {\rm K}
		\end{array} \right.
	\end{equation}	
	\begin{equation}
		k_{10}^H=\left\{
		\begin{array}{cl}
			3.6\times10^{-16}T_K^{3.640}{\rm exp}(\frac{6.035}{T_K})\;\;{\rm cm}^3{\rm s}^{-1} &\quad T_K \leq 10 {\rm K}\\
			3.1\times10^{-11}T_K^{0.357}{\rm exp}(-\frac{32}{T_K})\;\;{\rm cm}^3{\rm s}^{-1} &\quad T_K > 10 {\rm K}
		\end{array} \right.
	\end{equation}
These rates increase with temperature, which means that there is stronger collisional coupling for hotter gas than for cool gas. This effect is important if the gas was colder during the cosmic dawn due to unknown physics: the pre-reionization absorption trough might be shallower instead of steeper, in spite of having larger contrast of matter temperature from the CMB temperature. During EoR, the electron scattering  is more effective near the sources where there is partial ionization and high temperature. However, since we assume ionization fraction to be just the residual fraction outside of ionization bubbles, this effect is negligible in our work.

\section{Auto-Correlation of Dimensionless Brightness Temperature $\psi$} \label{autoh1sig} 
In our work, we wish to find the autocorrelation of $\psi$ (Eq.~(\ref{psidef})), which can be defined as, 
	\begin{eqnarray}
		\mu &=&\langle \psi_1\psi_2 \rangle-\langle \psi \rangle^2 \nonumber \\
			 &=& \langle n_1(1+\delta_1)(1-s_1)n_2(1+\delta_2)(1-s_2)\rangle-\langle n_1(1+\delta_1)(1-s_1)\rangle^2. \label{eq:defmu}
	\end{eqnarray}
Here, ($n_1$, $\delta_1$, $s_1$) and ($n_2$, $\delta_2$, $s_2$) are values of ionization, overdensity and heating ($T_\text{CMB}/T_S$) at point 1 (${\bf r_1}$) and at point 2 (${\bf r_2}$), respectively. Since the process of reionization is statistically homogeneous and isotropic, the autocorrelation function $\mu$ is function of $r=|{\bf r_2}-{\bf r_1}|$. To calculate $\mu$, we need to find all the pairs of points which are separated by a distance $r$, and take their weighed average over the entire space. To calculate probability of a pair with certain values, we use geometric arguments as described in RS18 and briefly summarized in  Appendix~{\ref{sec:Geometry}}.

In this paper, we assume that the density has no correlation with ionization or heating ($\langle n \delta\rangle = \langle s \delta\rangle = 0$)\footnote{These cross-correlations can be computed using excursion set formalism, e.g. \cite{FZH04a} for density-ionization cross-correlation. These terms are generally sub-dominant to other terms we retain (for discussion see e.g. RS18 and references therein).}. This gives us:
	\begin{align}
		\langle \psi_1\psi_2 \rangle &= (1+\xi)\langle n_1n_2(1-s_1)(1-s_2)\rangle, \nonumber \\
		\langle \psi\rangle &= \langle 1+\delta\rangle \langle n(1-s)\rangle = f_n -\langle n s\rangle \nonumber
	\end{align}
where, $\langle \delta \rangle=0$ and $\xi=\langle \delta({\bf r_1})\delta({\bf r_2}) \rangle$  is the autocorrelation function of the \ion{H}{1} density perturbation. We compute $\xi$ using the $\Lambda$CDM model power spectrum and assume the relative bias between the dark matter and the \ion{H}{1}, $b = 1$. $f_n=\langle n \rangle$ is defined as the average neutral volume fraction at that redshift. We also define, $\phi=n(1-s)$, to explore simplifying cases where we temporarily ignore the effect of density correlation $\xi$. In such case, two-point correlation function,
	\begin{align}
		\mu &= (1+\xi)\langle n_1n_2(1-s_1)(1-s_2)\rangle - \langle n(1-s)\rangle^2 \label{eq:mu_psi} \\
			&= (1+\xi)\langle \phi_1 \phi_2 \rangle - \langle \phi \rangle^2 \label{eq:mu_phi}
	\end{align}

The correlation functions  we calculate in RS18 and this paper are analytically derived. However, a function $\mu(r)$ is a valid correlation function only if it follows certain properties: a) It should be a finite positive value at $r=0$, $\mu(0)$; b) At any $r>0$, the value of correlation function $\mu(r)<\mu(0)$; c) Correlation should go to 0 at very large distance; d) The Fourier transform of the correlation function should be a positive definite function (power spectrum); e) Between $r=0$ and $r \to \infty$, the correlation function can be positive or negative, with the condition that its integration over all space, must be zero. We do not satisfy this  condition\footnote{As discussed in RS18, for computing ionization inhomogeneties we assume that the probability of finding an ionized region outside an ionization bubble is the global ionization fraction $f_i$.  While this is an excellent assumption for computing the correlation function on scales of interest to us, it violates the integral constraint on the  correlation function.}. 
However, we check for consistency of our formalism with the rest of the conditions.

We first discuss a few limiting cases (for details see RS18).  For scales greater than the largest bubbles, the fluctuations due to the ionization, heating, and Ly$\alpha$ coupling inhomogeneities vanish, and the \ion{H}{1} correlation function is determined by only density perturbations. In this limit we get:
	\begin{align}
		\mu &= \xi (f_n - \langle s\rangle)^2 \label{eq:corrls}.
	\end{align}
Here the correlation function scales as the density correlation function $\xi$. We also note that the correlation function at large scale vanishes when $f_n = \langle s \rangle$ (close to global heating transition). If the heating and Ly$\alpha$ coupling are uniform, the neutral gas of the IGM is at the same  spin temperature $T_{\rm bg}$. The correlation function simplifies in this case:
	\begin{align}
		\mu &= (1-s_b)^2((1+\xi)\langle n_1n_2\rangle- f_n^2) \label{eq:uniheat1}
	\end{align}
Here $s_b = T_{\rm CMB}/T_{\rm bg}$. If ionization fraction is very  small, $\langle n_1 n_2\rangle \simeq f_n^2 \simeq 1$. This gives us $\mu = \xi (1-s_b)^2$. Here the density fluctuations are enhanced by the temperature contrast between IGM and CMB. At late times $T_{\rm bg} \gg T_{\rm CMB}$ owing to X-ray heating, which drives $s_b$ to zero. This reduces the correlation function to the one dominated by density and ionization inhomogeneities:
	\begin{align}
		\mu = -f_n^2 + (1+\xi)\langle n_1 n_2\rangle 
	\end{align}
	\begin{align}
		\langle n_1n_2\rangle &= f_n- f_n \sum_{R_x}N(R_x)\frac{4\pi}{3}{R_x}^3 C(r,0,R_x,R_x)   \label{eq:corrneu_case1}
	\end{align}
This result is derived and explored in \cite{ZFH04} and RS18.

\subsection{Modelling and Notations} \label{sec:model_and_notations}
Our aim in this paper is to analytically model the correlation function of \ion{H}{1} brightness temperature fluctuations from the early phase of EoR owing to  scales that emerge due to ionization, heating and  Ly$\alpha$ coupling inhomogeneities. These inhomogeneities are caused by bubbles of a given size distribution, which evolves with time. These bubbles determine the scales of correlation.
The details of our modelling and main assumptions are discussed extensively in RS18. We briefly summarize them here for the sake of completeness, as only a subset of the details are given in this paper. 
We have assumed a topology where there are isolated, spherical self-ionized bubbles, surrounded by isotropic, smooth $T_S$ profiles which might overlap with one another and smoothly merge with the background. 
Given the statistical isotropy (we neglect redshift-space distortion) and homogeneity of the  process of  reionization and heating, our assumption of spherical bubbles and isotropic spin temperature profiles hold even though the individual bubbles might not be spherical.

We compute ionization and spin temperature autocorrelation and ionization-spin temperature cross correlation. We neglect, as noted above, the cross-correlation of density with ionization and spin temperature inhomogeneities as these contribute negligibly on the scales of interest. We neglect the clustering of self-ionized regions; a self-ionized region already accounts for clustering of ionizing sources at smaller scales. In RS18 we present cases that account for the clustering of self-ionized  region. This doesn't substantially  alter our main results, however, it might introduce new scales in the problem which correspond to the correlation scales of bubble centres. We note that this assumption is better for higher redshifts as the mean bubble separation is larger. 

\begin{figure}
	\centering
	\begin{minipage}{0.45\textwidth}
		\centering
		\includegraphics[width=0.9\textwidth]{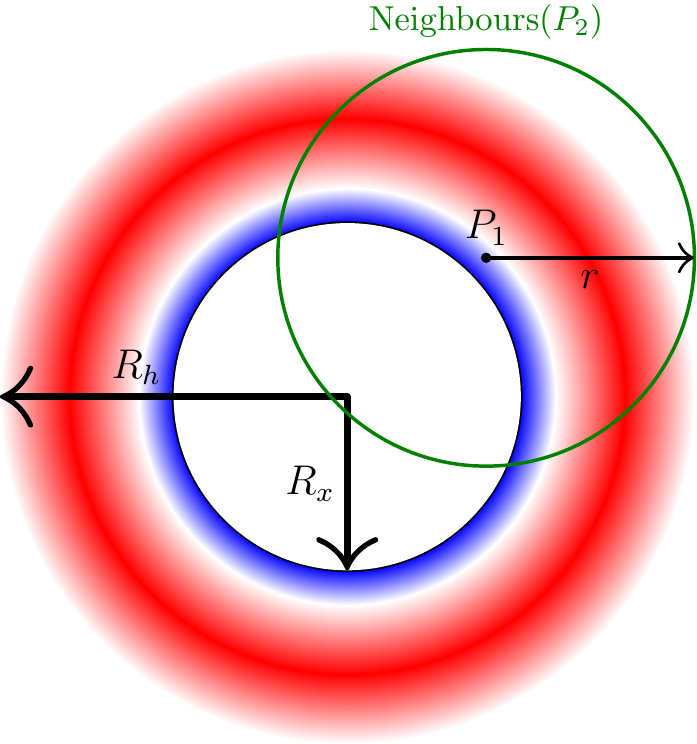}
		\caption{A cartoon for topology of the ionized region and its surrounding IGM is shown. The colour scheme shows the dimensionless brightness temperature, $\psi$. $\psi_i=0$ in the ionized region of size $R_x$ (with sharp boundary). It might be positive in neutral, heated and coupled region, and negative in neutral, non-heated and coupled region. The profile of radius $R_h$ merges smoothly with the background, which is not coupled in this case.} 
		\label{fig:Cartoon}
	\end{minipage}\hfill
\end{figure}

At any redshift, using excursion set formalism (Section~\ref{sec:photion}) and the matter power spectrum  given by $\Lambda$CDM model, we generate size distribution of self-ionized bubbles. Using Eqs.~(\ref{eq:tsbas}),~(\ref{eq:fintemp}),~(\ref{eq:lymanstar}), and~(\ref{eq:ycdef}), we calculate spin temperature in shells around these bubbles. The ionization volume fraction and volume fraction due to these shells is, respectively,
	\begin{align}
		f_i &= \sum_{R_x}\frac{4\pi}{3}N(R_x)R_x^3 \label{eq:fi} \\
		f_{hb} &= \sum_{R_x}\frac{4\pi}{3}N(R_x)(R_h^3-R_x^3) \label{eq:fhb}
	\end{align}
where $R_x$ are radii of ionization bubbles, $N(R_x)$ is the number density of bubbles with ionization radius $R_x$ and $R_h$ corresponds to the outer radius of the spin temperature profile for a given $R_x$ (Figure~\ref{fig:Cartoon}). During the initial phase, the ionization bubbles and surrounding temperature profiles are non-overlapping; we discuss the case of overlapping heating and Ly$\alpha$ coupled regions later in this section.

For any bubble of size $R_x$, we take shells of thickness $\Delta R(R_x, s)$ between radii $R_x$ and $R_h$, having temperature $s = T_{\rm CMB}/T_S$. A detailed description of notations followed in this paper is given in Table~1. To compute correlations we assume two random points separated by a distance $r$ as shown in Figure~\ref{fig:Cartoon}. The formalism used for the computation of correlation functions is described in the Appendix~{\ref{sec:comp}}. 

\subsection{Overlap} \label{sec:overlap_model}

At low redshift, spin temperature profiles around ionization bubbles are very large and overlap significantly. Thus, in such cases, $f_{hb}$ can be much larger than 1. In our previous paper (RS18), we had discussed a method to consistently take into account the overlap of heated regions. In this paper our principal aim is to model the fluctuation of Ly$\alpha$ radiation. Unlike the heated regions which merge after the heating transition, the mean free path of lower Lyman series photons is larger than the mean inter-bubble distance at all times for $z< 30$, the starting redshift  of our study. However, when the overlapped volume is very large ($f_{hb} \gg 1$), our formalism fails to generate a valid correlation function, as its Fourier transform (Appendix~\ref{sec:Geometry}) could  fluctuate and yield negative values  at large $k$. To avoid such unphysical results, we have taken a different approach to model overlaps in this paper. 

We calculate kinetic temperature and Ly$\alpha$ profiles up to very large distances, and start shedding outermost shells until $f_{hb}$ approaches 1. The energy and Ly$\alpha$ photons from excess  shells are uniformly distributed in the neutral universe (background as well as remaining shells). The bubbles are still likely to overlap due to randomness of their positions. To account for that, we use (Appendix~\ref{sec:overlap_maths}),
	\begin{align}
		f_h &\simeq \left(1+\frac{f_i}{f_{hb}}\right) (1-\mathrm{e}^{-f_{hb}}) -f_i \label{eq:fh} \\
		f_b &= 1-f_i-f_h \label{eq:fb}
	\end{align}
where $f_h$, corresponds to the  actual volume fraction occupied by heating bubbles (single counting of overlapped region) and $f_b$ is background fraction. $f_h$ approaches $f_{hb}$ when the $T_S$ profile volume fraction and ionization fraction are small. However, $f_h$ remains less than unity even if the value of $f_{hb}$ becomes much larger than unity. 

\startlongtable
\begin{deluxetable}{c|cc}
	\tablecaption{Notations \label{tab:table}}
	\tablehead{
		\colhead{Symbols} & \colhead{Explanation} 
	}
	\startdata
	$\delta$ & Over-density of \ion{H}{1} gas \\
	$n$ & Ionization state of \ion{H}{1} gas: Neutral point $n=1$ and ionized point $n=0$ \\
	$s$ & Temperature state defined as $s=T_{\rm CMB}/T_S$ \\
	$\psi$ & Dimensionless brightness temperature: $\psi=n(1+\delta)(1-s)$ \\	
	$\xi$ & Autocorrelation of overdensity $\delta$: $\xi =\langle \delta_1\delta_2 \rangle$\\
	$\phi$ & $\phi =n(1-s)$ \\
	$\mu$ & Autocorrelation of dimensionless brightness temperature $\psi$: $\mu =\langle \psi_1\psi_2 \rangle-\langle \psi \rangle^2$\\
	$f_i$ & Average ionized volume fraction \\
	$f_n$ & Average neutral volume fraction \\
	$f_{hb}$ & Total volume fraction due to $T_S$ profiles (without correcting for the overlaps)\\
	$f_h $ & Average $T_S$ profile volume fraction after correcting for the overlaps \\
	$f_b $ & Average background volume fraction \\
	$R_x$ & Radius of given ionization bubble\\
	$R_h$ & Outer radius of given $T_S$ profile: $R_h = R_h(R_x)$ \\
	$R_s$ & Inner radius of the shell with spin temperature $T_S=T_{\rm CMB}/s$ around given bubble \\
	$\Delta R_s$ & Thickness of the shell with spin temperature $T_S=T_{\rm CMB}/s$ around given bubble \\
	$N(R_x)$ & Number density of ionization bubbles of radius $R_x$ \\
	\enddata
\end{deluxetable}

\pagebreak
\subsection{Modelling Parameters}
In our analysis, we explore two parameters to model X-ray heating and  Ly$\alpha$ coupling: 
\begin{itemize}
  \item $N_\text{heat}$: Number of X-ray photons emitted per stellar baryon. For our study, we assume $N_{\rm heat}$ in the range: 0.1--10.0. For larger value of $N_{\rm heat}$, the heating is stronger, with higher values of $T_K$.
  \item $f_L$: Ratio of source luminosity of photons between Ly$\alpha$ and Lyman-limit to the luminosity of UV photons. We take $0.1<f_L<1000$ in this paper. As the value of $f_L$ increases, the coupling between $T_S$ and $T_K$ is stronger. This increases the absolute value of $\Delta T_B$ in both emission and absorption.
\end{itemize}
In this paper, we have not considered scenarios in  which the modelling parameters evolve with time.

\section{Results}\label{sec:res}

In the early phase of EoR, the brightness temperature fluctuations are determined by perturbations on  the scales  of heated and Ly$\alpha$ coupled regions. In the complete model, these regions have a range of sizes and have diffuse profiles (Appendix~\ref{sec:comp}). This makes it  difficult to disentangle the impact of input physics on the observable quantity. Therefore,   to  delineate the essential aspects of our formalism, we first  consider two simple models.

\subsection{A Simple Model: One bubble size, flat heating profile} \label{sec:1Z1L}
\begin{figure}
	\centering
	\begin{minipage}{0.45\textwidth}
		\centering
		\includegraphics[width=0.9\textwidth]{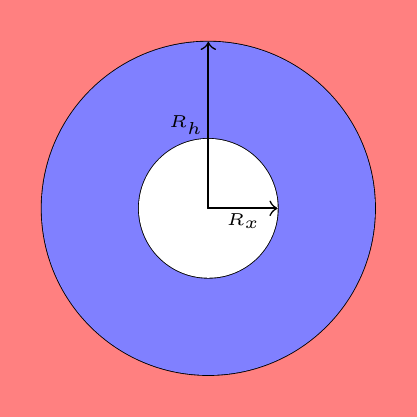}
		\caption{A simple case: an ionization bubble ($\psi_i =0 $) has one heated $T_S$ shells around it. Background is unheated, but coupled.}
		\label{fig:1shell}
	\end{minipage}\hfill
	\begin{minipage}{0.45\textwidth}
		\includegraphics[width=0.9\textwidth]{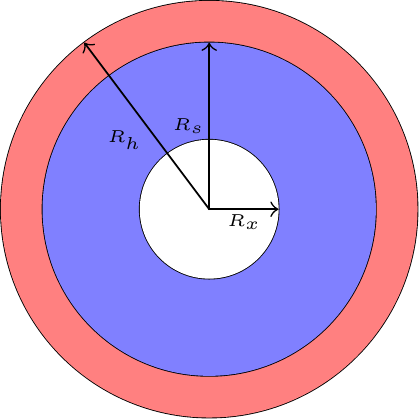}
		\caption{A simple case: an ionization bubble ($\psi_i =0 $) has two $T_S$ shells around it. The background is uncoupled ($\psi_b =0 $).}
		\label{fig:2shell}
	\end{minipage}
\end{figure}

We can study a simple model with single bubble size and a shell with uniform spin temperature around it (flat profile). There are small ionization bubbles embedded in larger heated bubbles (Figure~\ref{fig:1shell}). Ignoring density perturbations, there are only three values of $\phi=n(1-T_{\text{CMB}}/T_S)$ in this universe: $\phi_i=0$, $\phi_h= 1-T_{\text{CMB}}/T_{\text{heat}}$, and $\phi_b=1-T_{\text{CMB}}/T_{\text{bg}}$. When we include the density perturbations, the correlation function can be written as: 
	\begin{align}
		\mu	&= (1+\xi) \Big(\psi_h^2 (f_h - f_hf_b C(r, R_x, R_h, R_h) - f_if_h C(r,0,R_x,R_h) \nonumber \\
			& \quad \quad- f_i(1-f_i)(C(r,0,R_x,R_x)-C(r,0,R_x,R_h))) \nonumber \\
			&\quad + 2  \psi_h\psi_b f_h f_bC(r,R_x,R_h,R_h) \nonumber \\
			&\quad + \psi_b^2 f_b(1 - f_h C(r,R_x,R_h,R_h) - f_i C(r,0,R_x,R_h) ) \Big) - (\psi_b f_b + \psi_h f_h)^2 \label{eq:flatpro_fin}.
	\end{align}
where $R_x$ is the ionization bubble radius and $R_h$ is the heating bubble radius. Here, $C(x,P,Q,R)$ is a function with value between 0 and 1 (Appendix~\ref{sec:Geometry}). This result was derived and analysed in detail  for various limiting cases in RS18. This simple case does not allow for negative correlation since, within the bubble, the $\psi$ is positively correlated and it is not correlated outside  the bubble.

\subsection{A Simple Model: One bubble size, two shells, $\langle\psi\rangle=0$}\label{sec:1Z2L}
In the case where Ly$\alpha$ coupling is inhomogeneous, we can construct another simple scenario where there is only one ionized bubble size and the profile around the sources have two shells. The first shell is heated and coupled through Ly$\alpha$ radiation. The second shell is non-heated, but still coupled. The  region outside the second shell (background region) is neither heated nor coupled (Figure~\ref{fig:2shell}). This gives three values of $\phi=n(1-T_{\rm CMB}/T_S)$, when ignoring density fluctuations: $\phi_i=0$ in the  ionized regions; $\phi_1 =\phi> 0$, in the first shell; $\phi_2=-\phi<0$, in the second shell; and $\phi_b = 0$ in the background region. For simplicity we have assumed that the volume occupied by the two shells is the same, $f_1=f_2=f_h/2$ and $\phi_1 = - \phi_2 = \phi$. Therefore, $\langle\phi\rangle=f_1\phi_1+f_2\phi_2=0$.

From Figure~\ref{fig:2shell}, we can see that the correlation will be positive if both points are either in shell 1 or both in shell 2. If one point is in shell 1 and another in shell 2, then the correlation is negative. The two-point correlation function without density perturbation is,
	\begin{align}
		\mu &= \phi_1^2 P(\phi_1 \cap \phi_1) + \phi_2^2 P(\phi_2 \cap \phi_2) + 2\phi_1\phi_2 P(\phi_1 \cap \phi_2) \nonumber \\
			&= \phi^2 \Big(P(\phi_1) - P(\phi_1 \cap \phi_i) - P(\phi_1 \cap \phi_b) - P(\phi_1 \cap \phi_2) \nonumber \\
				& \qquad + P(\phi_2) - P(\phi_2 \cap \phi_i) - P(\phi_2 \cap \phi_b) - P(\phi_2 \cap \phi_1) - 2 P(\phi_1 \cap \phi_2) \Big), \nonumber
	\end{align}
which can be expanded using the same logic used in RS18 and Appendix~\ref{sec:comp},
	\begin{align}
		\frac{\mu}{\phi^2} &=  f_h (1 - C(r,0,R_x,R_h)) - f_i (1-f_i) [C(r,0,R_x,R_x)-C(r,0,R_x,R_h)] \nonumber \\  
				&\quad - (1-f_i) \frac{f_h}{2} [C(r,R_s,R_h,R_h) -C(r,R_x,R_s,R_h)]\nonumber \\
				&\quad - (1-f_i) f_h[C(r,R_x,R_s,R_s) + C(r,R_s,R_h,R_x) -C(r,R_s,R_h,R_s) -C(r,0,R_x,R_h) ].
				\label{eq:2shell_psi0}
	\end{align}
This results can also be obtained by simplifying the complete model (Eq~\ref{eq:psi12fin}) for approximations used in this section. At large scales, all the functions $C(.,.,.,.)$ tend to unity. In this case, Eq.~(\ref{eq:2shell_psi0}) approaches 0. This is expected, since at large scale ($r>2R_h$), the two-point correlation should vanish if we ignore density correlation. When $r=0$, $\mu=f_h\phi^2$. This is also as expected since, there is no probability of two points being in different shells at $r=0$. This expression takes negative value for a range of values of $r$, where the two points are more likely to be in different shells than in the same shell. 

\begin{figure}
	\centering
	\begin{minipage}{0.45\textwidth}
		\centering
		\includegraphics[width=0.9\textwidth]{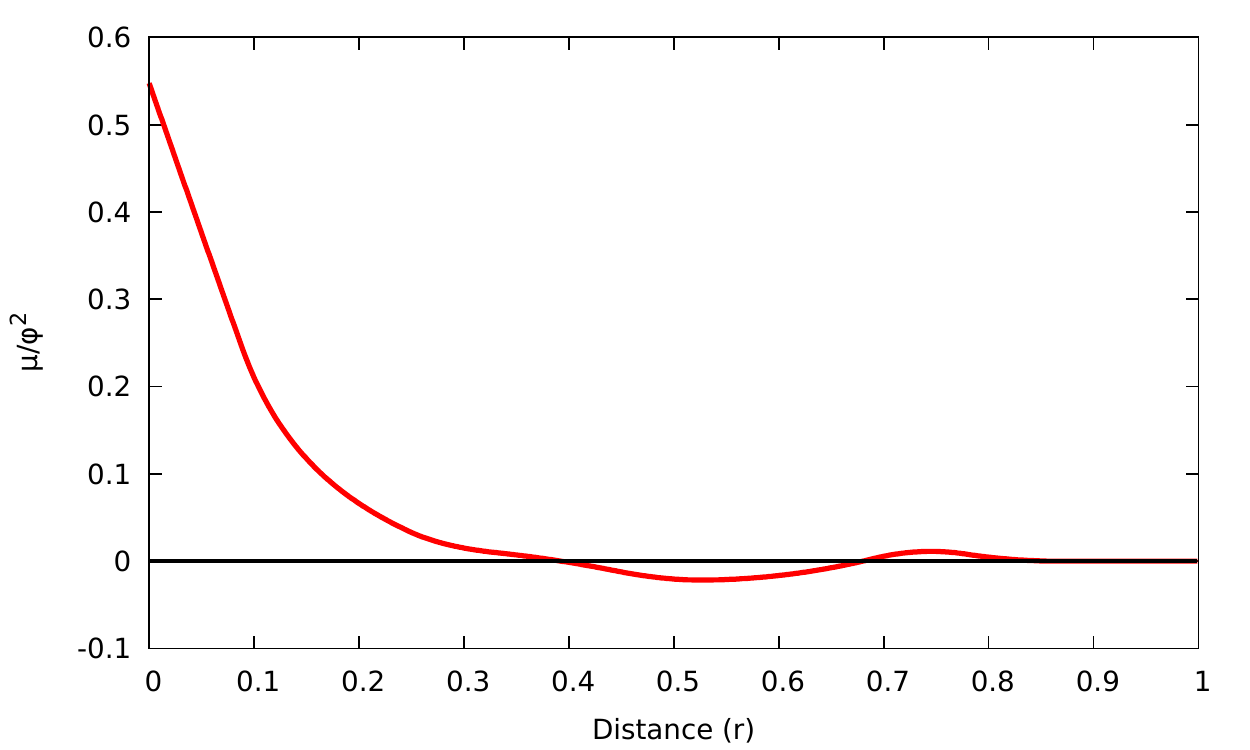}
		\caption{A case with negative correlation for certain scales is shown. Here, the ionization fraction $f_i = 0.01$ and heated fraction is $f_h = 0.55$. The distance scale would depend on the size of ionization bubble, which is chosen  to be 0.1 Mpc.} 
		\label{fig:2shell_corr}
	\end{minipage}\hfill
\centering
	\begin{minipage}{0.45\textwidth}
		\centering
		\includegraphics[width=0.9\textwidth]{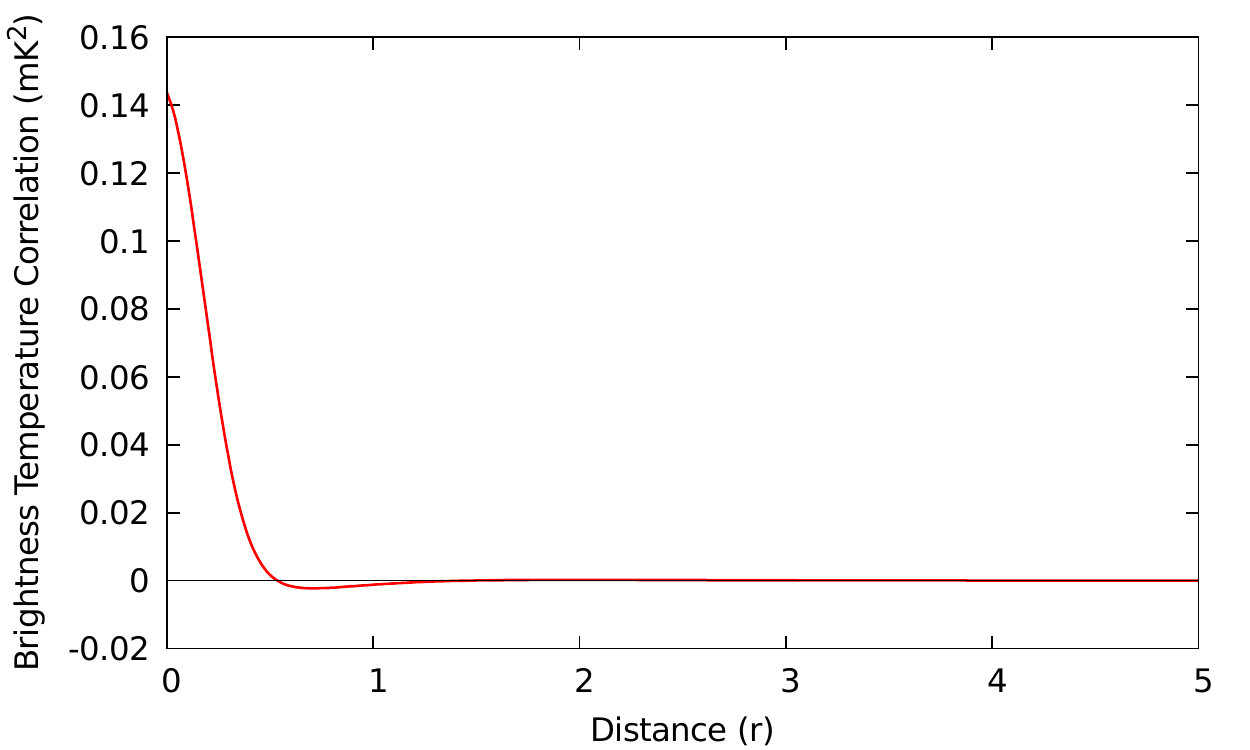}
		\caption{Neglecting density perturbations ($\xi =0$), the autocorrelation of \ion{H}{1} brightness temperature is shown at $z=20$ for the  complete model for  $\zeta=7.5$, $f_L=1.0$ and $N_{\rm heat}=1.0$. }
		\label{fig:corr_wd_20}
	\end{minipage}\hfill
\end{figure}

In Figure~\ref{fig:2shell_corr}, we have shown a case for this simple model where the autocorrelation of dimensionless brightness temperature is negative at certain scales. For scales $R_x+R_s < r < R_s+R_h$, depending on the values of various radii, two randomly chosen points can have higher probability of being in different shells than in the same shell, driving the correlation to negative value. For larger scales, $R_s+R_h < r < 2R_h$, both the points have finite probability of being in the outer shell, and zero probability of being in different shells of the same bubble, which leads the overall correlation at these scales to be positive. On scales $r > 2R_h$, the correlation function is zero. We note that  the possibility of negative correlation at any scale depends on values of $R_x$, $R_s$ and $R_h$ and their differences. It is entirely possible to have models where correlation function remains positive at all scales of interest.

For the complete model (discussed in detail in the next section and Appendix~\ref{sec:comp}), identifying the scales is more difficult since the ionization bubbles have a range of sizes, and the spin temperature profiles have a number of shells. However, if we ignore $\xi$, which is positively correlated at all scales of interest, we can still identify cases where the autocorrelation of $\phi$ goes negative. We show one such case in Figure~\ref{fig:corr_wd_20}. This constitutes the first important result of our analysis which underlines the importance of correlation analysis in real space  to extract the relevant physics. If correlation function is negative at certain scales, it will point to very specific geometry of the heated and Ly$\alpha$ coupled regions surrounding the self-ionized bubbles.

\pagebreak
\subsection{Complete model} \label{sec:res_comp}

In this paper, we have explored the correlation function and power spectrum evolution due to inhomogeneity of spin temperature using two modeling parameters: number of X-ray photons per stellar baryons $N_{\rm heat}$ and ratio of luminosity of Ly$\alpha$ photons to ionizing photons of the sources $f_L$. We have taken $\zeta \sim 7.5$ (Eq~\ref{eq:defzeta}), which is in agreement with the reionization optical depth ($\tau_{\rm reion} \simeq 0.055$) given by Planck (\cite{Planck2018}). We first present results for the $\Lambda$CDM model in the redshift range 10--30, without any additional cooling mechanism needed to explain the recent EDGES results (\cite{EDGES2018}). The results below redshift $z\simeq 12$ are not entirely reliable since, several approximation used in our formalism (including excursion set formalism) become less valid and eventually break down when the ionization volume fraction is large ($f_i> 0.1$) (\cite{2016MNRAS.457.1813F,2018MNRAS.473.2949G}). As noted above, to compute  brightness temperature mean and fluctuations, we have taken a size distribution of ionization bubbles, given by excursion set formalism and calculated spin temperature profiles surrounding these bubbles with a number of shells.

\begin{figure}
	\centering
	\begin{minipage}{0.45\textwidth}
		\centering
		\includegraphics[width=0.9\textwidth]{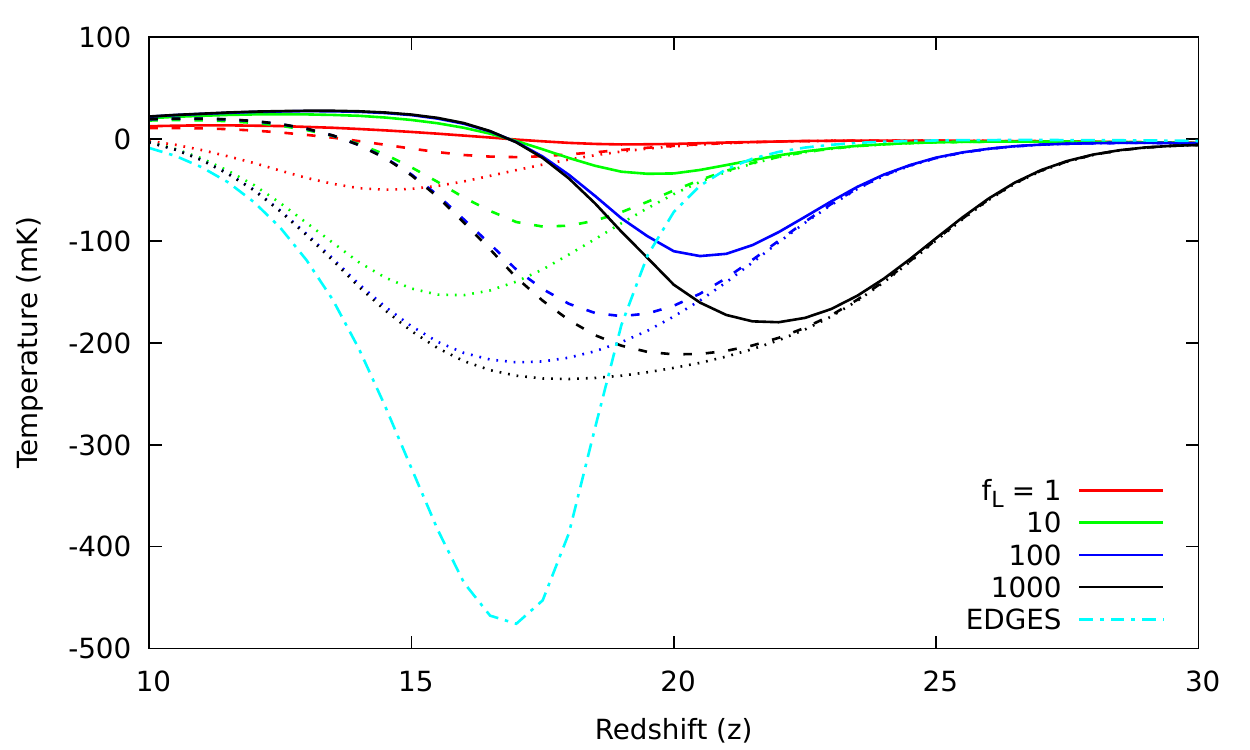}
		\caption{Global brightness temperature as function of redshift for various values of $f_L$ ranging from 0.1 to 1000 is shown. The solid lines are for $N_{\rm heat}=10$, long dashed lines for $N_{\rm heat}=1.0$ and short dashed lines for $N_{\rm heat}=0.1$. All plots have $\zeta=7.5$. The dot-dashed line represents a fiducial model that matches with EDGES observations.} 
		\label{fig:glob}
	\end{minipage}\hfill
\end{figure}

In Figure~\ref{fig:glob}, we show the evolution of global \ion{H}{1} brightness temperature as function of redshift for various combinations of modelling parameters. At $z\simeq 30$, the global signal starts  slightly negative as weak  collisional coupling drives the spin temperature towards matter temperature which is below CMB temperature (Eqs.~(\ref{eq:tsbas}) and~(\ref{overallnorm})). At smaller values of $z$ the behaviour is completely determined by the modelling parameters.  For higher values of $N_{\rm heat}$, the heating starts earlier and the absorption troughs are shallower. For higher value of $f_L$, the coupling starts earlier as well as stronger, and the overall strength of the signal is larger  at all redshifts before  complete coupling is achieved.  The redshift of heating transition (when the average gas temperature $T_K = T_{\rm CMB}$ and signal goes from absorption to emission) is only dependent on the heating ($N_{\rm heat}$) and  is independent of Ly$\alpha$ coupling ($f_L$). It happens sooner for higher value of $N_{\rm heat}$. 

\begin{figure}
	\centering
	\begin{minipage}{0.45\textwidth}
		\centering
		\includegraphics[width=1.0\textwidth]{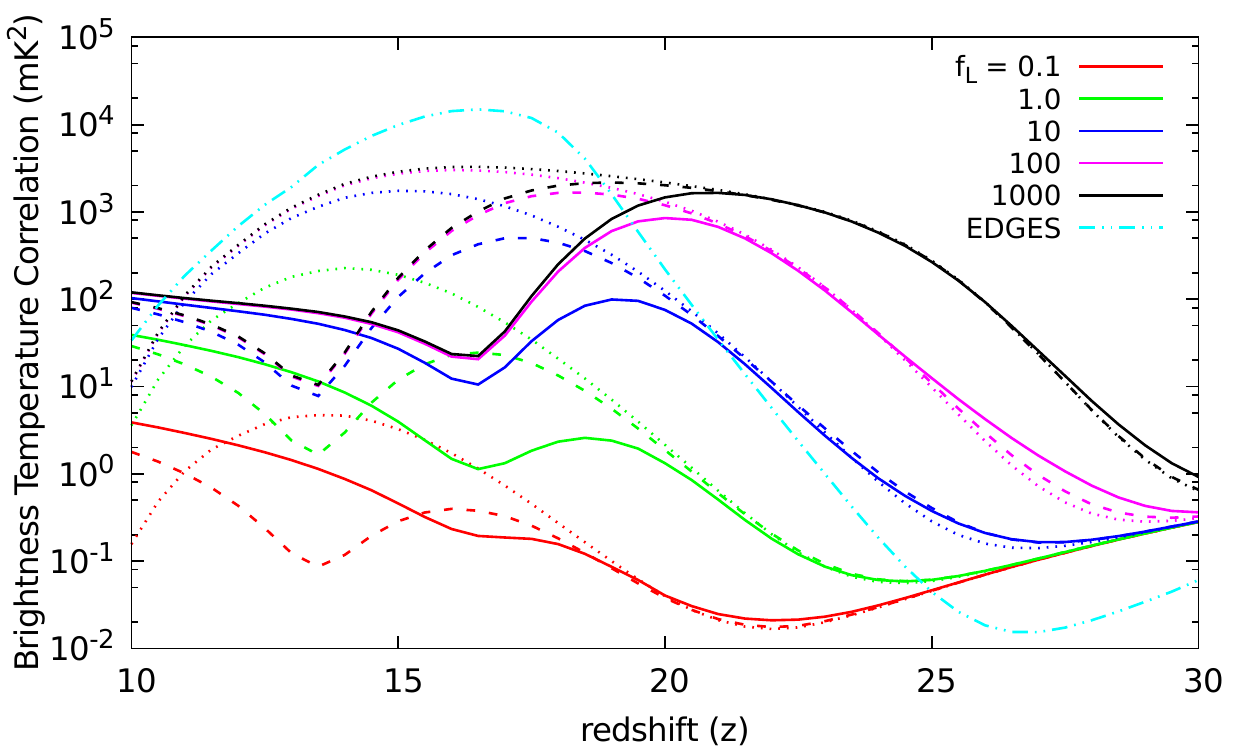}
	\end{minipage}\hfill
	\begin{minipage}{0.45\textwidth}
          \centering
		\includegraphics[width=1.0\textwidth]{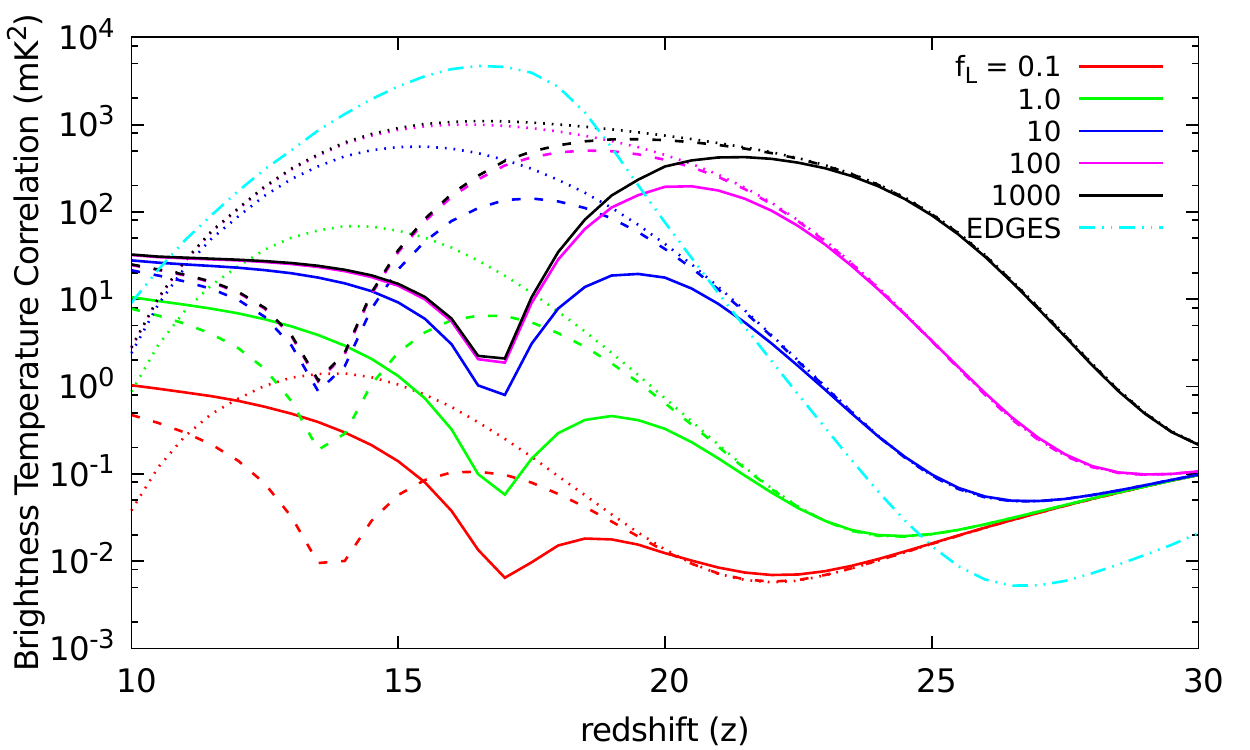}
	\end{minipage}
	\begin{minipage}{0.45\textwidth}
		\centering
		\includegraphics[width=1.0\textwidth]{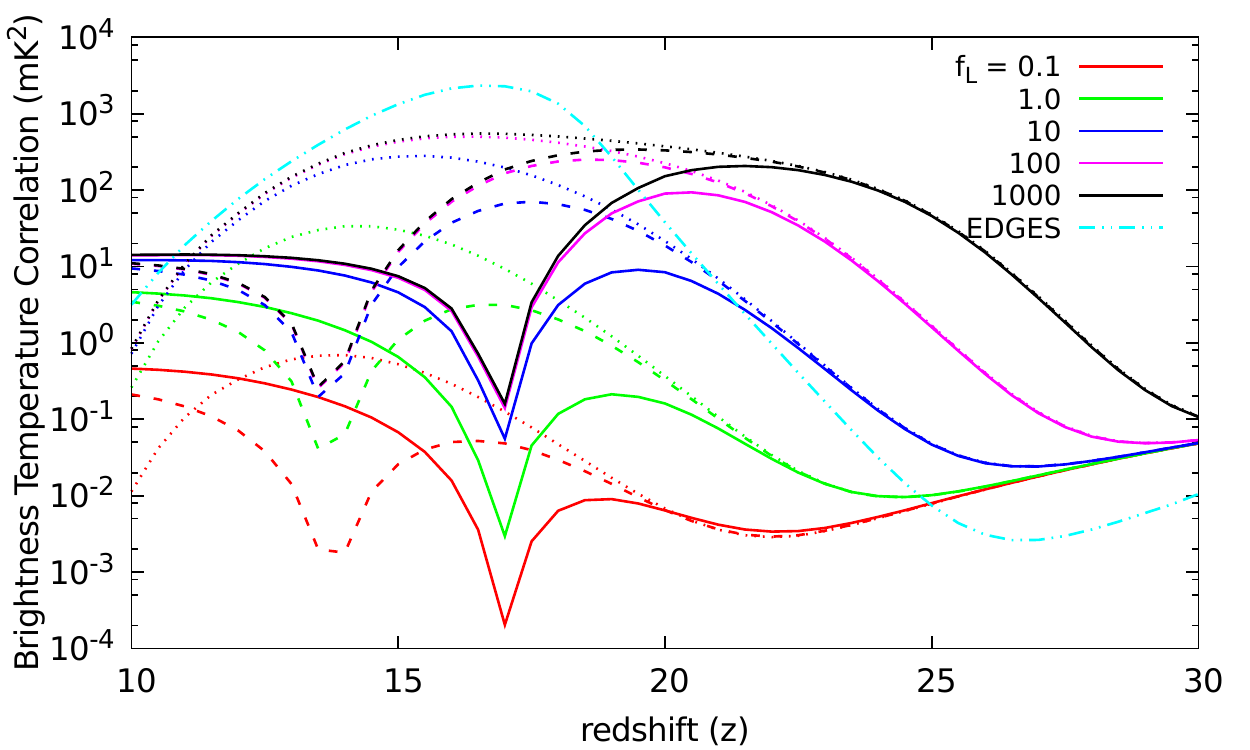}
	\end{minipage}\hfill
	\begin{minipage}{0.45\textwidth}
		\centering
		\includegraphics[width=1.0\textwidth]{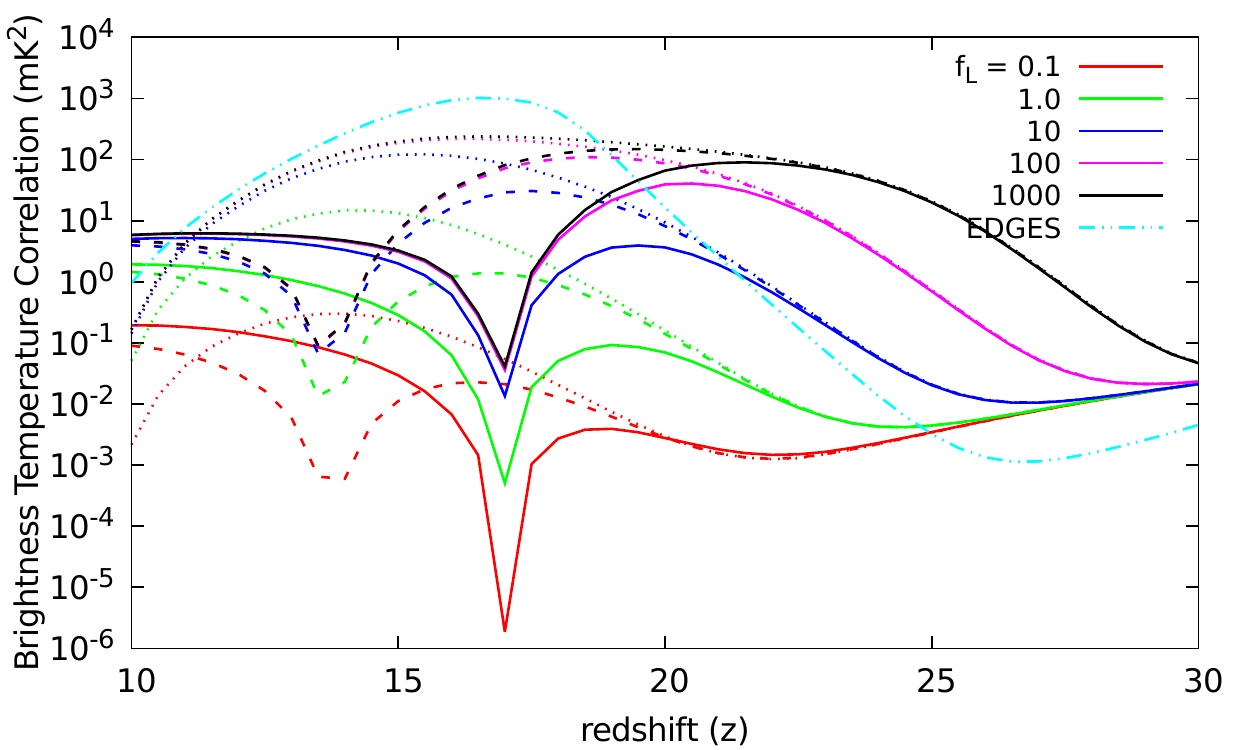}
	\end{minipage}
	\caption{Evolution of autocorrelation of \ion{H}{1} brightness temperature is shown for $r=0.5$~Mpc (top left panel),  $r=2.0$~Mpc (top right panel), $r=4$~Mpc (bottom left panel) and $r=8$~Mpc (bottom right panel) for $\zeta=7.5$, a range of $f_L$ varying from 0.1 to 1000 and three values of $N_{\rm heat}$: 10 (solid lines), 1.0 (long dashed lines) and 0.1 (short dashed lines). The dot-dashed lines represent a fiducial model that matches with EDGES observations.}
	\label{fig:corr}
\end{figure}

In Figure~\ref{fig:corr}, we show the evolution of correlation function at scales $r = 0.5\hbox{--}8 \, \rm  Mpc$ for different values of $N_{\rm heat}$ and $f_L$, using Eqs.~(\ref{overallnorm}),~(\ref{eq:fintemp}),~(\ref{eq:lymanstar}), and~(\ref{eq:psi12fin}). The correlation functions are large at small scales and decrease as the distance between two points increases. At very large scale, they approach Eq.~(\ref{eq:corrls}), where the density inhomogeneity is enhanced by the average of $(1-T_{\rm CMB}/T_S)$. On intermediate scales, the structure of correlation function is determined by the size distribution of bubbles and surrounding $T_S$ profiles.

As the main aim of this paper is to model brightness temperature fluctuations owing to incomplete Ly$\alpha$ coupling, we first discuss it qualitatively. As in the case of incomplete heating, the Ly$\alpha$ coupling can also be separated into near and far zone. In the near zone, the emission from a nearby source dominates the coupling. The strength of this coupling is determined by Eq.~(\ref{eq:lymanstar}), which shows that there always is a distance from the source at which complete Ly$\alpha$ coupling ($y_{\alpha,\star} T_K > T_{\rm CMB}$) can be established. Two conditions have to be met to ensure this creates a new length scale in the correlation function. First, this distance must exceed the size of the ionization region. As the coupling strength scales as $R_x^3$, larger ionization bubbles meet this requirement more easily. Second, the Ly$\alpha$ flux at this point from all the background sources must be smaller than the flux from the nearby source or we are in the limit of homogeneous coupling. The background intensity at any point is proportional to $R_{\rm max}(n)$ (Eq.~(\ref{eq:lyainf})) multiplied by the number density of sources at any redshift. In the initial phase, the background intensity is not high enough to cause complete Ly$\alpha$ coupling. During this phase, there are regions around individual sources that attain complete coupling and these regions are surrounded  by a background  in which only partial coupling has been attained. This creates inhomogeneities in Ly$\alpha$ coupling on the scales of these regions.  As the intensity builds in the background owing to the birth of new self-ionized regions and it  reaches levels sufficient to cause complete coupling, these inhomogeneities disappear. The nature of these inhomogeneities is determined by $f_L$, the atomic structure of neutral hydrogen, and the excursion set formalism.

At $z\simeq 30$, only the  collisional coupling is effective. As it is weak and the Ly$\alpha$ intensity is small, all the curves shown in the figure start with small values  of correlation function with similar strength. The fluctuations are dominated by density perturbations  as number density of ionizing sources is small and  the ionization, heating, Ly$\alpha$ coupled  fractions are tiny.  For small $f_L$, the correlation declines with times as the collisional coupling  weakens owing to the fall of  the number density of particles  with the expansion of the universe. This situation is only reversed  when  Ly$\alpha$ coupling becomes efficient. For higher value of $f_L$, this coupling occurs sooner, leading to an increase in   fluctuations with time until complete coupling is achieved. After this period, the fluctuations are determined by heating and ionization  inhomogeneities. The position of the first peak in Figure~ \ref{fig:corr} depends strongly on the heating parameter as the peak is determined by the thermal  evolution of the  background; the correlation only start decreasing when the background is sufficiently heated. This phase is described in detail in RS18. 

The correlation function approaches zero  near the heating transition. At large scales, the  signal vanishes completely when $T_{\rm CMB}/T_K$ approaches $f_n$ (Eq.~(\ref{eq:corrls})). We again see the effect of $N_{\rm heat}$ on the redshift of heating transition. Inhomogeneous collisional coupling and shape of the temperature profiles, which are determined by spectrum of X-ray photons ($\alpha$ and $\nu_{\rm min}$, see RS18), can potentially change the redshift and depth of heating transition by a small amount. We do not study this effect in the current paper. After the heating transition, the effect of inhomogeneous $T_S$ decreases and main source of fluctuations is ionization inhomogeneity. Their effect is suppressed if the heating or coupling is not saturated (very small values of $N_{\rm heat}$ and $f_L$). In general, larger $N_{\rm heat}$ creates larger heating profiles and causes correlation at larger scales. However, these profiles also merge sooner and wipe out heating fluctuations at those scales.

\begin{figure}
	\centering
	\begin{minipage}{0.45\textwidth}
		\centering
		\includegraphics[width=1.0\textwidth]{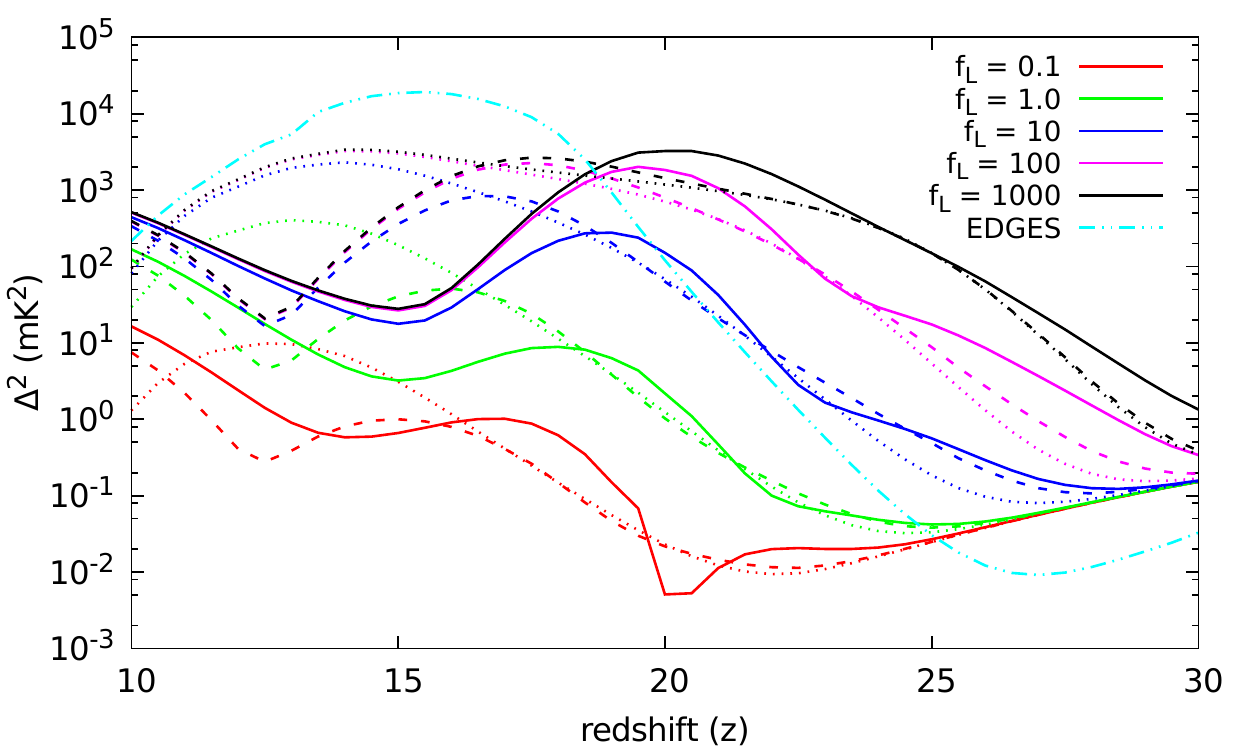}
	\end{minipage}\hfill
	\begin{minipage}{0.45\textwidth}
		\includegraphics[width=1.0\textwidth]{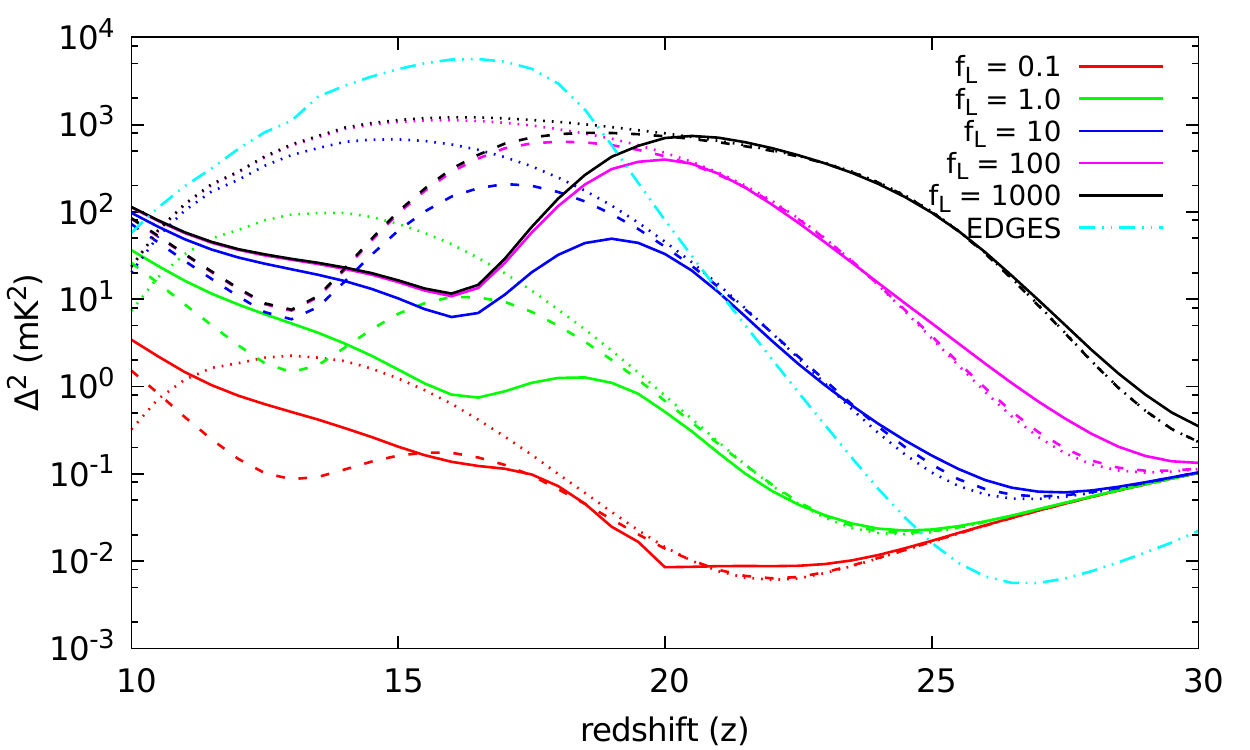}
	\end{minipage}
        \begin{minipage}{0.45\textwidth}
		\centering
		\includegraphics[width=1.0\textwidth]{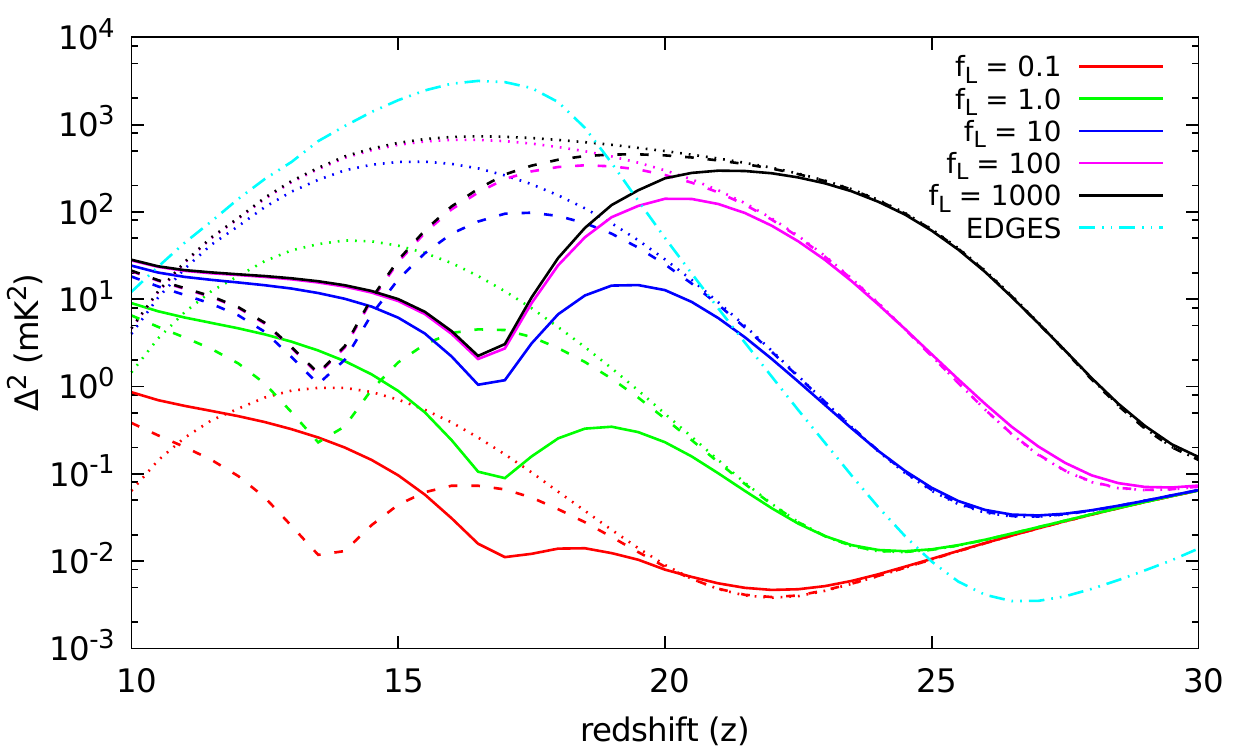}
	\end{minipage}\hfill
	\begin{minipage}{0.45\textwidth}
		\includegraphics[width=1.0\textwidth]{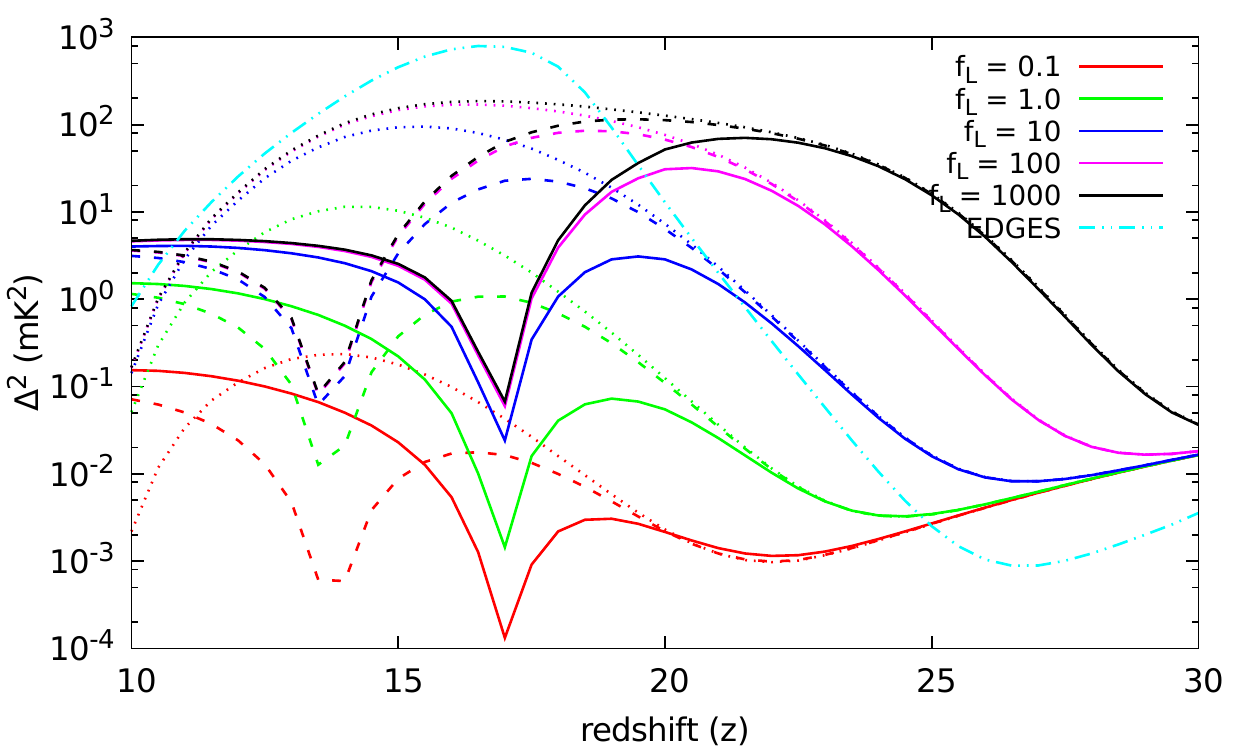}
	\end{minipage}
	\caption{Evolution of $\Delta^2 = k^3P(k)/2 \pi^2$ of \ion{H}{1} brightness temperature is shown for $k=2 \, \rm Mpc^{-1}$ (top left panel), $k=1 \, \rm Mpc^{-1}$ (top right panel), $k=0.5 \, \rm Mpc^{-1}$ (bottom left panel) and $k=0.125 \, \rm Mpc^{-1}$ (bottom right panel)  for $\zeta=7.5$, a range of $f_L$ varying from 0.1 to 1000 and three values of $N_{\rm heat}$: 10 (solid lines), 1.0 (long dashed lines) and 0.1 (short dashed lines). The dot-dashed lines represent a fiducial model that matches with EDGES observations.}
	\label{fig:PS}
\end{figure}

The \ion{H}{1} signal from the epoch of cosmic dawn and reionization has been extensively studied in the literature using semi-analytic methods and large-scale simulations (e.g. \cite{2007MNRAS.376.1680P,2008ApJ...689....1S,2010A&A...523A...4B,2010MNRAS.406.2421S,2012Natur.487...70V,2013MNRAS.435.3001T,2013MNRAS.431..621M,2014MNRAS.443..678P,2015PhRvL.114j1303F,2015MNRAS.447.1806G,2016MNRAS.459.2342M,2017MNRAS.464.3498F,2017MNRAS.468.3785R}).
Our formalism has been built on geometric arguments, which are intuitively easier to visualize in real space. The entire correlation structure can be written in terms of a single polynomial function: $C(.,.,.,.)$ (section~\ref{sec:Geometry}). Taking Fourier transform with respect to  the first argument $r$ of this function yields the power spectrum. In Figure~\ref{fig:PS}, we show evolution of power spectrum $\Delta^2 = k^3P(k)/2 \pi^2$ for  a range of  $k$. These figures show similar evolutionary trend as the correlation functions (Figure~\ref{fig:corr}). 

While comparing our results with simulations, we have focused on three features: (a) the number of  peaks in the power spectrum, (b) the  amplitude  $\Delta^2(k)$ for a range of scales $k \sim 0.1\hbox{--}0.5 {\rm Mpc}^{-1}$ and (c) the difference between the redshift of heating transition and the redshift of power spectrum minimum.

Existing results in the literature show that, for $k \simeq 0.1\hbox{--}0.5 \, \rm Mpc^{-1}$, there are generally two or three peaks of power spectrum as a function of redshift (\cite{2008ApJ...689....1S,2010A&A...523A...4B,2013MNRAS.431..621M,2015MNRAS.447.1806G,2016MNRAS.459.2342M,2017MNRAS.464.3498F}).
At high redshifts, when the Ly$\alpha$ coupling and X-ray heating commence, they  create fluctuations of $T_S$ in the medium. If fluctuations in these two fields dominate at widely different times, there will be two distinct peaks at high redshift: one due to coupling inhomogeneities and the other (generally at lower redshift than former) due to heating inhomogeneities (\cite{2008ApJ...684...18C,2007MNRAS.376.1680P,2015aska.confE...3A}).  After the heating transition, there is a third, smaller peak at low redshifts, when the power spectrum is dominated by ionization inhomogeneities (e.g. \cite{2007MNRAS.376.1680P,LateHeat2,2015MNRAS.447.1806G}). We studied the 
two peaks owing to heating and ionization inhomogeneities in RS18. 

\begin{figure}
\centering
\begin{minipage}{0.45\textwidth}
		\centering
		\includegraphics[width=0.9\textwidth]{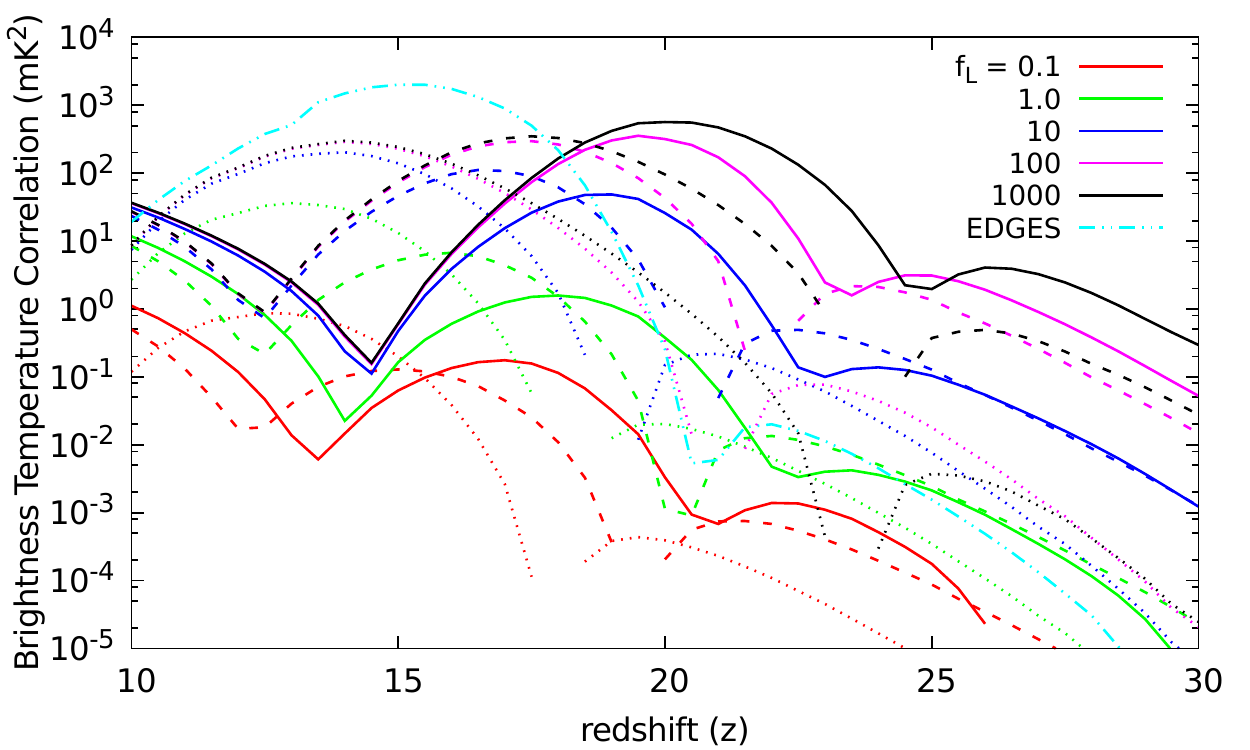}
		\caption{Neglecting density perturbations, the evolution of autocorrelation of \ion{H}{1} brightness temperature is shown for $r=0.5$~Mpc   for $\zeta=7.5$, a range of $f_L$ varying from 0.1 to 1000 and three values of $N_{\rm heat}$: 10 (solid lines), 1.0 (long dashed curves) and 0.1 (short dashed lines). The dot-dashed line represents a fiducial model that matches with EDGES observations.}
		\label{fig:corr_wd}
	\end{minipage}\hfill
\end{figure}

The third peak owing to the inhomogeneous Ly$\alpha$ coupling is expected at early times for the following reason: as large self-ionizing regions are born these fluctuations should initially build and then diminish as the contrast between Ly$\alpha$ coupling in the near and far zone reduces, finally disappearing when complete Ly$\alpha$ coupling is established. 
We see this additional peak in Figure~\ref{fig:corr_wd} in which the impact of density perturbations in neglected. However, in the complete model, we find a weak peak owing to this effect only at small scales in the power spectrum ($k = 2 \, \rm Mpc^{-1}$ in Figure~\ref{fig:PS}). Generally this possible  additional peak is masked by  density perturbations. The strength  of this peak can be understood in terms of the evolution of number density  of large self-ionized regions at early times and influence region of Ly$\alpha$ photons. For  $z < 30$, the number density of self-ionizing bubble builds exponentially in the excursion set formalism. While this creates inhomogeneities owing to geometry seen in Figure~\ref{fig:2shell}, it also causes a rapid build up of the background Ly$\alpha$ photons, rapidly destroying  the contrast between the near and far zone. At any point the background flux gets nearly equal contribution from sources within  the (comoving) radius $R_{\rm max}(n)$  (Eq.~(\ref{eq:lyainf})). This radius is close to  600~Mpc for $n=2$ at $z\simeq 25$. This large influence region contributes to wiping out  the contrast in a short span. We note that $R_{\rm max}(n)$ (for small principal quantum numbers  $n>2$) determines the length scale at which the physics needs to be captured to study the Ly$\alpha$ generated inhomogeneities. It might  be  difficult to achieve  it using an $N$-body simulation as the box size is generally smaller than this length scale.
In our results for large $N_{\rm heat}$ and small $f_L$, we get three peaks at large $k$ (top left panel of Figure~\ref{fig:PS}), since the heating, and by extension collisional coupling inhomogeneities dominate \textit{before} the Ly$\alpha$ inhomogeneities commence.

While $\Delta^2(k)$  agrees with the  results of simulations for smaller 
scales, for $k\simeq 0.1 \, \rm Mpc^{-1}$, our results give  less power as  compared to   simulations (bottom right panel of Figure~\ref{fig:PS}). Given the small sizes of ionization bubbles at high redshifts, the only contribution to fluctuations at $k \sim 0.1 {\rm Mpc}^{-1}$ is due to density and spin temperature fluctuations. For the models of heating and Ly$\alpha$ coupling used in this paper,  the contribution due to heating and coupling is dominated by far-away sources which diminishes fluctuations at large scales. Therefore at these scales, the spin temperature inhomogeneities are negligible and only the density fluctuations are enhanced by the average contrast of \ion{H}{1} spin temperature with CMB temperature.  While it is conceivable that the higher power at large scales in simulation is owing to finite box size (for a discussion see \cite{2011MNRAS.414..727Z,2015MNRAS.447.1806G}), which might not allow one to take into account the contribution  of far-away sources whose impact   tends to homogenize the fluctuations of $T_S$ at large scales, a more detailed comparison with simulations is hard as the parameter  range used  is generally not the same.

When the average spin temperature equals to the CMB temperature during the heating transition, the power spectrum at small $k$   reaches a minimum value (Figure~\ref{fig:PS}).
However, for larger $k$ (small $r$), even during the heating transition there are significant fluctuations due to inhomogeneities of spin temperature and ionization which delays the minima of the power spectrum. 
Figure~\ref{fig:PS} shows that, for $k = 2 {\rm Mpc}^{-1}$, the minima of power spectrum depends on the value of $f_L$ even though the heating transition is  independent of it. 
In general, the minimum of power spectrum occurs during or after global heating transition, depending on scales and modelling parameters. This is in agreement with simulations that explore the dependence of these inhomogeneities on modelling parameters (\cite{2013MNRAS.431..621M,2015MNRAS.447.1806G,2016MNRAS.459.2342M,2017MNRAS.464.3498F}).

\subsection{Implication of EDGES detection}

Recent EDGES observation (\cite{EDGES2018}) reported  a sky-averaged absorption feature of strength $\Delta T \simeq -500 \,\rm mK$ in the frequency range $70$--$90$~MHz, corresponding  to a redshift range $15$--$19$ for the redshifted \ion{H}{1} line. It can be shown that for standard recombination and thermal history the minimum temperature of the gas at $z\simeq 17$ is $T_K \simeq 6 \, \rm K$. It follows from Eqs.~(\ref{eq:tsbas}) and~(\ref{overallnorm}) that the absorption trough should not have been deeper than $-180 \, \rm mK$. 

One possible explanation of the EDGES result is there is additional  radio background in the redshift range at $15 < z< 19$; in this case we can replace $T_{\rm CMB}$ with the $T_{\rm CMB} +T_{\rm radio}$ in Eq.~(\ref{eq:tsbas}) in the relevant  redshift range (\cite{feng18}, \cite{2018ApJ...868...63E}, \cite{sharma18}). With this replacement and suitable choice of $T_{\rm radio}$ we can re-derive all our results of this paper for compatibility with EDGES result.


Another plausible explanation invokes the additional cooling of baryon owing to interaction between dark matter and baryons \footnote{It might be possible to  distinguish this scenario from the one invoking a higher radio background if the radio background leave signatures of its characteristic fluctuations due to radio sources. We hope to return to this issue in a future work.}.  In this case, we can explain the EDGES detection using  Eqs.~(\ref{eq:tsbas}) and~(\ref{overallnorm}) if: ({\it a}) Ly$\alpha$ photons globally couple the spin temperature to matter temperature, i.e. $T_K y_\alpha \gg T_{CMB}$, such that $T_S = T_K$ at $z\simeq 19$, and ({\it b}) $T_K  \simeq 2.5 \, \rm K$ \footnote{EDGES detection implies a sharp trough in the signal at $z\simeq 19$ and an equally sharp rise at $z\simeq 15$. As the noise level for the detection is $\simeq 20 \, \rm mK$, the drop at higher redshift can arise from complete Ly$\alpha$ coupling being established close to $z\simeq 19$ with the  rapid heating being responsible for the sharp rise at smaller redshift. It should be noted that one of the implications of the EDGES results is that $f_L$ cannot be too large, otherwise the complete Ly$\alpha$ coupling would be established at a higher redshift and even in the absence of additional cooling the signal would be close to $-200 \, \rm mK$ at $z > 19$ which should be observable but is not seen by EDGES.}. We consider this theoretical extension to explain EDGES result in this paper\footnote{Other possible explanations of this result include a possible systematic error \citep{hills18} and  absorption  from spinning dust grains in the  Galactic ISM  \citep{draine-miralda} among others.}.

We consider a fiducial model which roughly fits the EDGES results. We assume dark matter-baryon interaction of the form described in \cite{Barkana2018}, with cross section $\sigma_1=5 \times 10^{-24}\; {\rm cm}^2$ and the ratio of dark matter to proton mass $m_{dm}/m_p = 0.001$. Such an interaction helps cool the baryon gas temperature sufficiently to explain the EDGES result. For $f_L = 2$ and $N_{\rm heat} =0.08$, we  get an absorption trough in the global signal similar to the one observed by EDGES data (Figure~\ref{fig:glob}). In Figures~\ref{fig:corr} and~\ref{fig:PS}, we show the correlation functions and power spectrum for a range of scales, taking into account our fiducial model to replicate the EDGES results. 

The main impact of EDGES observation on the expected correlations is to boost the signal by nearly an order of magnitude in the redshift range $15 < z < 19$, even as compared to the most optimistic models\footnote{We have not incorporated the enhancement in the signal due to the inhomogeneous velocity-dependent cooling of gas within this model (\cite{fialkov18, 2018PhRvL.121l1301M}).} (low $N_{\rm heat}$ and high $f_L$) in the usual case \footnote{We notice a decrease in the signal in this case at $z\simeq 30$. This model-dependent decrement is owing to cooler baryons causing  a decrease in the efficiency of coupling from collisions.}. This is entirely owing to a decrease of $T_S$ which boosts  $s$ in Eq.~(\ref{psidef}). The correlation function scales roughly as $(1-s)^2$.  Even though this result has been arrived at within the framework of a model involving baryon cooling, this result holds  when excess radio background is responsible for the deep absorption feature as $s$ increases by a similar factor in that case too. 

This result is also robust to change in other modelling parameters $f_L$ and $N_{\rm heat}$. 
Therefore, an important  prediction of the EDGES detection  is that  the deeper absorption trough in the global signal gives a  corresponding increase in the fluctuating component  too. Even when the spin temperature field is very uniform, with  Ly$\alpha$-coupled and unheated gas, the low  spin temperature would enhance the underlying density inhomogeneities (Eq~\ref{eq:uniheat1}). As we have shown in Section~\ref{sec:1Z2L}, the spin temperature field can be negatively correlated at some scales. In such scenarios, the fluctuating component of the signal could be less than given by (Eq~\ref{eq:uniheat1}). We tried  to produce such models for parameters needed to explain the EDGES data and found it very difficult to anti-correlate the spin temperature field. Hence we infer that the minimum value of correlation function which correspond to the EDGES signal is  given by (Eq~\ref{eq:uniheat1}),  using the temperature derived from the trough in the global signal. If the global signal has a trough of $\sim 500$ mK at $z\simeq 17$, then assuming complete coupling and no heating (no fluctuations due to these two fields), the auto correlation function at $r=2$ Mpc and $r=4 $ Mpc should be $\sim 5100\; ({\rm mK})^2$ and $\sim 2500\; ({\rm mK})^2$ respectively. 

Many currently  operational  (e.g. LOFAR,  MWA, PAPER, GMRT) and upcoming radio interferometers (HERA, SKA)  have the capability to detect the fluctuating component of the \ion{H}{1} signal in the redshift range $8 <z < 25$ (for details e.g. \cite{2015aska.confE...3A,2015aska.confE...1K,2014MNRAS.439.3262M}). 
We discuss the detectability of this signal especially in light of the recent EDGES results. These radio interferometers directly  measure visibilities and their correlations which can be related to  the power spectrum of the \ion{H}{1} signal (e.g. \cite{Bharadwaj01,ZFH04}). This data analysis  can readily be  extended to  the image plane (which is a byproduct of the analysis pipeline e.g. \cite{2017ApJ...838...65P} for LOFAR). This means   real-space correlation functions can also be  used for computation of the signal (e.g. \cite{2008ApJ...673....1S}).  
As an approximate rule, one could use  $r \simeq \pi/k$ to  shift   from Fourier to real space.  

SKA1-LOW is expected to detect the \ion{H}{1} signal at $z\simeq 16$ with a signal-to-noise varying  from 100 to 10  for $0.1 < k < 0.6 \, \rm Mpc^{-1}$ for a signal strength $\Delta^2(k) \simeq 10^2 \, \rm (mK)^2$ (for details see e.g. \cite{2015aska.confE...1K}). EDGES results predict a signal strength nearly a factor of 5 larger which means the signal-to-noise would be significantly higher (Figure~\ref{fig:PS}). The ongoing experiment LOFAR's best  upper limit correspond to $(80 \, \rm mK)^2$ ($k \simeq 0.05 \, \rm Mpc^{-1}$) in the redshift range $9.6 < z < 10.6$ in 10~hours of integration \footnote{Angular scale above which the \ion{H}{1} signal can be reliably measured for most  ongoing and upcoming radio interferometers is  a few arcminutes;  $1'$ corresponds to nearly 3~Mpc (comoving) at $z \simeq 15$ or  these telescopes are sensitive to linear scales larger than  $5\hbox{--}10 \, \rm  Mpc$ (comoving). However, these telescopes have frequency resolution which correspond to  much smaller linear scales, e.g. MWA's frequency resolution of 40~kHZ corresponds to nearly 1~Mpc (comoving) along the  line of sight. Or the 3-d \ion{H}{1} signal is probed with different resolution on the sky plane as compared to the line of sight.}. LOFAR has the  frequency range  to probe the redshift range  of   EDGES detection. If the  noise  properties  at smaller frequencies
($\simeq 80 \, \rm MHz$) behave roughly as the one observed at higher frequencies ($\simeq 110 \, \rm MHz$), LOFAR might be able to detect the signal in a few hundred hours of integration. 

SKA1-LOW (and SKA2) have the capability to detect the EoR signal in the redshift range $20 < z < 25$. A signal of  $100 \, \rm (mK)^2$ at $z \simeq 25$ is detectable by a deep SKA1-LOW survey  with signal-to-noise of five  for $k < 0.1 \, \rm Mpc^{-1}$ (\cite{2015aska.confE...1K}). However, we notice that EDGES result places significant constraint on the signal in this era. This, as noted above, is owing to the fact that EDGES results imply a smaller value of $f_L$ which results in  smaller signal for $z > 19$ as compared to models with larger $f_L$ which give significantly higher signal (Figure~\ref{fig:PS}). 

\section{Summary and conclusions} \label{sec:sumcon}

The main aim of this paper is to extend the analytic formalism presented in our previous work (\cite{RS18}) to explore the epoch of partial  Ly$\alpha$  coupling. Following RS18, we generate the size   distribution of self-ionized regions  using excursion set formalism for $\Lambda$CDM model. Around these bubbles, we create spin temperature profiles due to X-ray heating and Ly$\alpha$ coupling, which merge smoothly with a uniform background. 
We model these spin temperature $T_S$ profiles using two parameters: $N_{\rm heat}$, the number of X-ray photons per stellar baryons, and $f_L$, the ratio of Ly$\alpha$ to ionizing photons. Our analytic formulation allows us to explore relevant  physical processes and study  both their individual as well as combined  impact on the \ion{H}{1} signal. 

We study  the evolution of correlation properties of the \ion{H}{1} in the redshift range $10 \le z \le 30$ for many possible scenarios, with greater focus on higher redshifts at which partial Ly$\alpha$ coupling  plays an important role (Figures~\ref{fig:2shell_corr}, \ref{fig:corr_wd}, \ref{fig:corr},~\ref{fig:PS}). We find reasonable agreement with existing semi-analytic and $N$-body simulation results. As we compute correlation functions in both real and Fourier space, we find a possible case where the correlation function in real space is negative owing to partial heating and Ly$\alpha$ coupling (Figure~\ref{fig:2shell_corr}). 

We also analyse the implications of the recent  EDGES detection of the global \ion{H}{1} signal in the redshift range $15 <z < 19$. Generically, EDGES detection results in higher correlation signal in the redshift range of the  detection but lower signal  at higher redshifts, as compared to the most optimistic models which do not take into account this detection (Figures~\ref{fig:corr} and \ref{fig:PS}).  

In this paper,  we study the implications of  the standard $\Lambda$CDM model. The resultant size distribution of self-ionized regions   and hence the fluctuation scales of 
brightness temperature would be different for an extension of this model, which  we hope to explore in the future work (e.g. \cite{2009JCAP...11..021S,2016JCAP...04..012S}).

Understanding  theoretically the \ion{H}{1} signal from EoR/cosmic dawn remains a challenge. Given the large amount of uncertainty in the physics of ionizing  sources, IGM, feedback mechanisms, etc, it is important to explore a wide set of modelling paradigms. While $N$-body simulations are important to understand and image the \ion{H}{1} field, analytic methods, like the one presented in this paper, are suited to predict the statistical quantities like correlation function and power spectrum. Since our formalism is not limited by the size of the simulation box, we can easily incorporate a variety of physical processes at very small or very large scale, e.g. the influence region of Ly$\alpha$ photons. Also, this formalism is computationally cheaper, which means we can explore a large set of modelling parameters and their degeneracies at a fraction of computation resources taken by $N$-body simulation. Since $N$-body simulations, semi-analytic, and analytic formalisms each have their own set of assumptions, strengths, and weaknesses, it is beneficial to apply all  these methods to unravel the complex physics of reionization

\acknowledgments

The authors would like to thank Akash Kumar Patwa for valuable discussions and insights.

\appendix
\section{Probability}
	\begin{align}
		P(A|B)&=\frac{P(A\cap B)}{P(B)} \label{eq:B2} \\
		P((A\cap B)|C) &=P(A|(B\cap C))\; P(B|C) \label{eq:B3} \\
		P(A\cap B) &=P(A)-P(A\cap \tilde B) \label{eq:AandB} 
	\end{align}
	
\section{Geometry} \label{sec:Geometry}
If point 1 is located between distance $P$ and $Q$ $(P<Q)$ from the center of a sphere, then $C(x,P,Q,R)$ is the probability that its neighbour point 2 at distance $x$ from point 1 is located outside the concentric sphere of radius $R$.
	\begin{equation}
		C(x,P,Q,R)=\left\{
		\begin{array}{cl}
			0 &\quad x \leq R-Q\\
			1 &\quad x \leq P-R\\
			1 &\quad x \geq R+Q\\
			\frac{1}{2} \frac{Q^3-(R-x)^3}{Q^3-P^3} + \frac{3}{8x} \frac{Q^2-(R-x)^2}{Q^3-P^3} \left[\frac{Q^2+(R-x)^2}{2}+(x^2-R^2)\right] & R-Q \leq x \leq R-P,\; x > Q-R\\
			\frac{Q^3-R^3}{Q^3-P^3} &R-Q \leq x \leq R-P,\; x \leq Q-R\\
			\frac{1}{2} \frac{2Q^3-(R+x)^3-P^3}{Q^3-P^3} + \frac{3}{8x} \frac{(R+x)^2-P^2}{Q^3-P^3} \left[\frac{(R+x)^2+P^2}{2}+(x^2-R^2)\right] & |P-R| \leq x \leq Q-R,\;x < P+R\\
			1 - \frac{R^3}{Q^3-P^3} & P+R \leq x \leq Q-R,\;\\
			\frac{1}{2} \frac{(x-R)^3+Q^3-2P^3}{Q^3-P^3} + \frac{3}{8x} \frac{Q^2-(x-R)^2}{Q^3-P^3} \left[\frac{Q^2+(x-R)^2}{2}+(x^2-R^2)\right] & R+P \leq x \leq R+Q,\;x > Q-R\\
			\frac{1}{2} + \frac{3}{8x} \frac{P+Q}{P^2+PQ+Q^2} \left[\frac{P^2+Q^2}{2}+(x^2-R^2)\right] & R-P \leq x \leq P+R,\; x \geq Q-R
		\end{array} \right.
	\end{equation}
Three dimensional fourier transform of $1-C(x,P,Q,R)$ is useful while calculating power spectrum.
	\begin{align}
		FT(1-C(x,P,Q,R)) &= \frac{12 \pi}{k^6 (Q^3 - P^3)} \nonumber \\
		&\quad ([kQ\;{\rm Cos}(kQ)-{\rm Sin}(kQ)]-[kP\;{\rm Cos}(kP)-{\rm Sin}(kP)])[kR\;{\rm Cos}(kR)-{\rm Sin}(kR)]\nonumber 
	\end{align}
	
\section{Overlap}\label{sec:overlap_maths}
Assume that there is a sphere with two shells: $R_i$ is the radius of inner shell (actually a sphere), and $R_o$ is the outer shell radius. We wish to randomly put $N$ such spheres in a large box of volume $V$ ($V \gg 4\pi/3\;R_o^3$), in such a way that the inner shells of any two spheres must not overlap with each other, but they can overlap with outer shell of another sphere. Outer shells of two or more spheres can overlap with one another. We wish to calculate the total volume fraction occupied by the outer shells of these $N$ spheres.

When $N=1$, there is only one sphere randomly placed within the box. The fraction of volume of this box occupied by the inner and outer shells is, respectively,
	\begin{align}
		g_i = \frac{4 \pi}{3} \frac{R_i^3}{V} \quad \text{and} \quad g_o = \frac{4 \pi}{3} \frac{R_o^3-R_i^3}{V} \nonumber
	\end{align}
with  the total volume fraction occupied by the sphere, $g_t = g_i + g_o$. Here we are only interested in the volume fraction occupied by the outer shells. Thus, we call this fraction (for $N=1$), $g_1=g_o$. For $N=2$, we need to randomly place the second sphere where the first sphere is already placed in the box. Across multiple experiments, where the second sphere is placed randomly, the average total volume fraction that is occupied by the outer shell of these two sphere is, 
	\begin{align}
		g_2 = g_1 + g_1 (1-g_t) = g_1 + g_1 (1-(g_1+g_i)), \nonumber
	\end{align}
where the first term $g_1$ is due to first sphere and second term is due to second sphere which has ($1-g_t$) probability of overlap with the first sphere. Placing a  third sphere in this scenario ($N=3$), 
	\begin{align}
		g_3 &= g_2 + g_1 (1-(2g_i + g_2)), \nonumber
	\end{align}	
where the first term $g_2$ is due to first two sphere and second term is due to the third sphere which can overlap with these two spheres. This can be expressed recursively for $N$ spheres as,
	\begin{align}
		g_N &= g_{N-1} + g_1 (1-((N-1)g_i + g_{N-1})) \nonumber \\
			&= g_{N-1} (1-g_1) + g_1 (1-(N-1)g_i) \nonumber \\
			&= \sum_{k=0}^{N-1} g_1 (1-g_1)^k (1-(N-1-k)g_i) \nonumber \\
			&= \left(1+\frac{g_i}{g_1}\right)(1-(1-g_1)^N)-Ng_i. \label{eq:overlap1}
	\end{align}	
Here we can define, $f_i = Ng_i$, the total volume fraction occupied by the inner shells of all spheres (since they do not overlap, it is simple addition), and $f_{ob} = Ng_o$, the total sum of fraction occupied by the outer shells independently, including the multiple counting of overlapped part. The actual volume fraction occupied by the outer shells is, $f_o = g_N$ In the limit where $g_1 \ll 1$ and $N \gg 1$, 
	\begin{align}
		1-(1-g_1)^N \simeq 1-{\rm exp} (-Ng_1). \nonumber
	\end{align}	
Thus, Eq~\ref{eq:overlap1} can be written as, 
	\begin{align}
		f_o = \left(1+\frac{f_i}{f_{ob}}\right)(1-{\rm exp} (-f_{ob}))-f_i. \label{eq:overlap2}
	\end{align}	
The results match very well with the simulation where a large number of spheres are randomly arranged within a box with above conditions.

\section{Complete Model $(\mu=\langle \psi_1 \psi_2\rangle-\langle \psi \rangle^2)$}\label{sec:comp}
In this section, we develop a formalism to calculate correlation of dimension-less brightness temperature $\psi$ for epochs at which ionization volume fraction is small. This ensures that ionization bubbles are separate and non-overlapping. However, as described in section~\ref{sec:overlap_model} and Appendix~\ref{sec:overlap_maths}, our formulation allows us to deal with  overlap of heating bubbles. 

Since we have already assumed that the  cross correlation of density with ionization or spin temperature is negligible, we try to find the autocorrelation of $\phi=n(1-s)$ (henceforth referred to as `temperature' in this section) (Eq~\ref{eq:mu_phi}). Therefore to calculate correlation at scale $r$, we need to find pairs of points such that $\phi_1 \neq 0$ and $\phi_2 \neq 0$, where $\phi_1$ and $\phi_2$ are temperatures of point 1 and 2 which are separated by distance $r$.
	\begin{align}
		\langle \phi_1 \phi_2\rangle &= \phi_b^2 P(\{\phi_1=\phi_b\} \cap \{\phi_2=\phi_b\}) + 2 \phi_b \sum_{R_x} \sum_{\phi (R_x)} \phi P(\{\phi_1=\phi_b\} \cap \{\phi_2=\phi\}) \nonumber \\
				&\quad +  \sum_{R_x} \sum_{\phi_p (R_x)}\sum_{R'_x} \sum_{\phi_q (R'_x)} \phi_p \phi_q P(\{\phi_1=\phi_p\} \cap \{\phi_2=\phi_q\}) \label{eq:psi12def}
	\end{align}
Here, $\phi_i$ and $\phi_b$ are the temperatures of ionized region and background regions respectively. $P(\{\phi_1=\phi_p\} \cap \{\phi_2 = \phi_q \})$ is the joint probability of point 1 having temperature $\phi_p$ and point 2 having temperature $\phi_q$. In above equation, first, second and third terms correspond respectively to (1) both points being in background region, (2) one point being in background and another in a heated bubble, and (3) both points being in some heated bubble. If one or both points are in ionized region, the term corresponding to that pair will be 0. When both points are in background we use Eq~\ref{eq:AandB}, 
	\begin{align}
		P(\{\phi_1=\phi_b\} \cap \{\phi_2=\phi_b\}) &= P(\phi_1=\phi_b) - P(\{\phi_1=\phi_b\} \cap \{\phi_2 \neq \phi_b\}) \nonumber \\
			&= P(\phi_1=\phi_b) - P(\{\phi_1=\phi_b\} \cap \{\phi_2=\phi_i\}) -  \sum_{R_x} \sum_{\phi (R_x)} P(\{\phi_1=\phi_b\} \cap \{\phi_2=\phi\}) \label{eq:psibb}
	\end{align}
In the case where both points are partially heated, these points can be within the same bubble or different bubbles. Within the same bubble, they can be in the same shell ($\phi_1=\phi_2$) or in different shells. This gives,
	\begin{align}
		\sum_{R_x} \sum_{\phi_p (R_x)}\sum_{R'_x} \sum_{\phi_q (R'_x)} \!\!\! \phi_p \phi_q P(\{\phi_1=\phi_p\} \cap \{\phi_2=\phi_q\}) 
			&= \sum_{R_x} \sum_{\phi_p (R_x)} \phi_p^2 P(\{\phi_1=\phi_p\} \cap \{\phi_2=\phi_p\} \cap \{\text{same}\}) \nonumber \\
				& + \sum_{R_x} \sum_{\phi_p (R_x)} \sum_{\phi_q (R_x) \neq \phi_p} \phi_p \phi_q P(\{\phi_1=\phi_p\} \cap \{\phi_2=\phi_q\} \cap \{\text{same}\}) \nonumber \\
				& + \sum_{R_x} \sum_{\phi_p (R_x)}\sum_{R'_x} \sum_{\phi_q (R'_x)}\!\! \phi_p \phi_q P(\{\phi_1=\phi_p\} \cap \{\phi_2=\phi_q\} \cap \{\text{diff}\}) \label{eq:psipqdef}
	\end{align}
$P(\{\phi_1=\phi_p\} \cap \{\phi_2=\phi_p\} \cap \{\text{same}\})$ is the probability that both points are in the same bubble with the same temperature $\phi_p$. This is straightforward to calculate. However, since we allow overlap of heated profiles, the simplest derivation is not necessarily correct. Instead, we expand it further,
	\begin{align}
		\sum_{R_x} \sum_{\phi_p (R_x)} \!\! \phi_p^2 P(\{\phi_1=\phi_p\} \cap \{\phi_2=\phi_p\} \cap \{\text{same}\}) &= \sum_{R_x} \sum_{\phi_p (R_x)} \!\! \phi_p^2 (P(\phi_1=\phi_p)-P(\{\phi_1=\phi_p\} \cap \{\phi_2 \neq \phi_p\}) \nonumber \\
			&= \sum_{R_x} \sum_{\phi_p (R_x)} \!\! \phi_p^2 \Big(P(\phi_1=\phi_p) - P(\{\phi_1=\phi_p\} \cap \{\phi_2 = \phi_b\})\nonumber \\
			&\qquad\qquad\qquad\qquad\qquad\qquad - P(\{\phi_1=\phi_p\} \cap \{\phi_2 = \phi_i\})\Big)\nonumber \\
			&\; - \sum_{R_x} \sum_{\phi_p (R_x)} \sum_{\phi_q (R_x) \neq \phi_p} \!\! \phi_p^2 P(\{\phi_1=\phi_p\} \cap \{\phi_2 = \phi_q\} \cap \{\text{same}\}) \nonumber \\
			&\; - \sum_{R_x} \sum_{\phi_p (R_x)} \sum_{R'_x}\sum_{\phi_q (R'_x)} \!\! \phi_p^2 P(\{\phi_1=\phi_p\} \cap \{\phi_2 = \phi_q\} \cap \{\text{diff}\})\label{eq:psippdef}
	\end{align}

Putting Eq~{\ref{eq:psibb}, \ref{eq:psipqdef} and \ref{eq:psippdef}} in Eq~\ref{eq:psi12def}, we get,
	\begin{align}
		\langle \phi_1 \phi_2\rangle&= \phi_b^2 P(\phi_1=\phi_b) - \phi_b^2 P(\{\phi_1=\phi_b\} \cap \{\phi_2=\phi_i\}) \nonumber \\
			&\quad + \sum_{R_x} \sum_{\phi_p (R_x)} \!\! \phi_p^2 P(\phi_1=\phi_p) -\sum_{R_x} \sum_{\phi_p (R_x)} \!\! \phi_p^2 P(\{\phi_1=\phi_p\} \cap \{\phi_2 = \phi_i\}) \nonumber \\
			&\quad - \sum_{R_x} \sum_{\phi (R_x)} (\phi_b-\phi)^2 P(\{\phi_1=\phi_b\} \cap \{\phi_2=\phi\}) \nonumber \\
			&\quad + \sum_{R_x} \sum_{\phi_p (R_x)} \sum_{\phi_q (R_x)} \phi_p (\phi_q-\phi_p) P(\{\phi_1=\phi_p\} \cap \{\phi_2=\phi_q\} \cap \{\text{same}\}) \nonumber \\
			&\quad + \sum_{R_x} \sum_{\phi_p (R_x)}\sum_{R'_x} \sum_{\phi_q (R'_x)} \phi_p (\phi_q-\phi_p) P(\{\phi_1=\phi_p\} \cap \{\phi_2=\phi_q\} \cap \{\text{diff}\}) \label{eq:psi12exp} 
	\end{align}

Now, we calculate each term separately. $P(\phi_1=\phi_b)=f_b$ is the probability of a point being in background region. $P(\phi_1=\phi_p)= 4 \pi/3\; N(R_x) f_b ((R_{s_p}+\Delta R_{s_p})^3-R_{s_p}^3)$ is the probability of a point having temperature $\phi_p$ in a bubble of ionization radius $R_x$.

The probability of one point being in background and second point being ionized can be calculated by assuming that point 1 is in ionized region of radius $R_x$, and calculating the probability that its neighbour point 2, at distance $r$ is outside temperature profile of that bubble ($C(r, 0, R_x, R_h)$). Given this condition, the probability of point 2 being in background region is equal to the average background fraction $f_b$. Thus, 
 \[ P(\{\phi_1=\phi_b\} \cap \{\phi_2=\phi_i\})= f_b \sum_{R_x}N(R_x)\frac{4\pi}{3} R_x^3 C(r,0,R_x,R_h) \]

When point 1 has temperature $\phi_p$ and point 2 is ionized, they both can be in the same bubble or in different bubbles, which respectively give first and second terms on the right hand side of the following equation.
	\begin{align}
		\sum_{R_x} \sum_{\phi_p (R_x)} \!\! \phi_p^2 P(\{\phi_1=\phi_p\} \cap \{\phi_2 = \phi_i\})
			&= \sum_{R_x} \sum_{\phi_p (R_x)} \!\! \phi_p^2  \frac{4\pi}{3}N(R_x)\bigg({R_x}^3(1-f_i)\nonumber \\
			&\qquad\qquad\qquad[C(r,0,R_x,R_{s_p})-C(r,0,R_x,R_{s_p}+\Delta R_{s_p})] \nonumber \\  
			&\qquad\qquad+ f_b((R_{s_p}+\Delta R_{s_p})^3-R_{s_p}^3) \sum_{R'_x}\frac{4\pi}{3}N(R'_x){R'_x}^3  C(r,0,R'_x,R'_h)\bigg) \label{eq:psipi}
	\end{align}
	
When point 1 is in background and point 2 has temperature $\phi$, we can expand it as,
	\begin{align}
		P(\{\phi_1=\phi_b\} \cap \{\phi_2=\phi\}) &= P( \phi_1(\text{out other})|(\phi_1(\text{out same})\cap\{\phi_2=\phi\})) P(\phi_1(\text{out same}) \cap \{\phi_2=\phi\}), \nonumber
	\end{align}
where the first term on right hand side, $P(\phi_1(\text{out other})|(\phi_1(\text{out same})\cap \{\phi_2=\phi\} ))$, is the probability that point 1 is in background region given that {point 2 is partially heated with temperature $\phi$ and point 1 is not inside the same bubble in which point 2 is}. This probability equals the fraction of the universe heated to background temperature, $f_b$. The second term on right hand side, $P(\phi_1(\text{out same})\cap \{\phi_2=\phi\})$  is the probability that point 1 is out of the bubble in which point 2 is, and point 2 is partially heated with temperature $\phi$. As point 2 can be in bubble with any ionization radius $R_x$, $P(\phi_1(\text{out same})|\{\phi_2=\phi\}(R_x))$ is the probability that point 1 is out of the bubble which has ionization radius $R_x$ and which contains point 2 with temperature $\phi$. This equals the probability that point 1 is out of the bubble with outer radius $R_h$ in which point 2 is located between radius $R_s$ and $R_s+\Delta R_s$.
	\begin{align}
		\sum_{R_x} \sum_{\phi (R_x)} P(\{\phi_1=\phi_b\} \cap \{\phi_2=\phi\}) &= f_b \sum_{R_x}N(R_x)\frac{4\pi}{3}\sum_{\phi(R_x)} f_b((R_s+\Delta R_s)^3-R_s^3) C(r,R_s,R_s+\Delta R_s,R_h) \label{eq:psibp}
	\end{align}

$P(\{\phi_1=\phi_p\} \cap \{\phi_2=\phi_q\} \cap \{\text{diff}\})$ gives the  probability that point 1 has $\phi = \phi_p$, point 2 has $\phi=\phi_q$ and they both belong to different bubbles. We take a simple assumption that if point 2 is outside the bubble in which point 1 is, then its probability of having $\phi=\phi_q$ is equal to the global probability of $\phi_q$ temperature shell. Here, either $p$ or $q$ could have been point 1. Therefore,
	\begin{alignat}{3}
		&& \sum_{R_x} \sum_{\phi_p (R_x)}\sum_{R'_x} \sum_{\phi_q (R'_x)}\!\! \phi_p (\phi_q-\phi_p) & \nonumber \\
			&& \quad\quad P(\{\phi_1=\phi_p\} \cap \{\phi_2=\phi_q\} \cap \{\text{diff}\}) &= \sum_{R_x} N(R_x)\frac{4\pi}{3} \sum_{\phi_p (R_x)} f_b((R_{s_p}+\Delta R_{s_p})^3-R_{s_p}^3) \nonumber \\
			&& &\quad \sum_{R'_x}N(R'_x)\frac{4\pi}{3} \sum_{\phi_q (R'_x)} f_b((R'_{s_q}+\Delta R'_{s_q})^3-{R'_{s_q}}^3) \phi_p (\phi_q-\phi_p) \nonumber \\
 			&& &\quad  \frac{C(r,R_{s_p},R_{s_p}+\Delta R_{s_p},R_h)+C(r,R'_{s_q},R'_{s_q}+\Delta R'_{s_q},R'_h)}{2} \label{eq:psipqdif}
	\end{alignat}
$P(\{\phi_1=\phi_p\} \cap \{\phi_2=\phi_q\} \cap \{\text{same}\})$ gives the  probability that point 1 has $\phi = \phi_p$, point 2 has $\phi=\phi_q \neq \phi_p$ and they both belong to the same bubble. If point 1 is located at distance between $R_{s_p}$ and $R_{s_p}+\Delta R_{s_p}$ from the centre of the sphere, then fraction of its neighbours at distance $r$ which are outside the sphere of radius $R_{s_q}$ and inside sphere of radius $R_{s_q}+\Delta R_{s_q}$ can be computed. However, since bubbles can overlap, point 2 can be neutral or ionized, which leads to:  
	\begin{align}
		\sum_{R_x} \sum_{\phi_p (R_x)} \sum_{\phi_q (R_x) \neq \phi_p}\!\!\!\! \phi_p \phi_q P(\{\phi_1=\phi_p\} \cap (\phi_2=\phi_q) \cap (\text{same})) &= \sum_{R_x} N(R_x)\frac{4\pi}{3} \sum_{\phi_p (R_x)} f_b((R_{s_p}+\Delta R_{s_p})^3-R_{s_p}^3) \nonumber \\
			&\quad \sum_{\phi_q (R_x) \neq \phi_p} \phi_p \phi_q 	(1-f_i) \Big( C(r,R_{s_p},R_{s_p}+\Delta R_{s_p},R_{s_q})  \nonumber \\
			&\qquad\qquad -C(r,R_{s_p},R_{s_p}+\Delta R_{s_p},R_{s_q}+\Delta R_{s_q}) \Big) \label{eq:psipqsame}
	\end{align}
Putting all equations together, including the influence of $\xi$ and simplifying in terms of $F(x,P,Q,R)=1-C(x,P,Q,R)$, we get,
	\begin{align}
		\mu &= \xi(f_b\phi_b + \langle \phi_h \rangle )^2 + (1+\xi)\Bigg((\phi_b^2 f_b + \langle \phi_h^2 \rangle )\sum_{R_x}\frac{4\pi}{3}N(R_x) R_x^3 F(r,0,R_x,R_h) \nonumber \\
			& \qquad + f_b \sum_{R_x} \frac{4\pi}{3}N(R_x) \sum_{\phi (R_x)} (\phi_b-\phi)^2 f_b((R_s+\Delta R_s)^3-R_s^3) F(r,R_s,R_s+\Delta R_s,R_h) \nonumber \\
			&\qquad + \sum_{R_x}\frac{4\pi}{3} N(R_x) {R_x}^3 \sum_{\phi (R_x)} \!\! \phi^2 (1-f_i) [F(r,0,R_x,R_s)-F(r,0,R_x,R_s+\Delta R_s)] \nonumber \\  
			&\qquad -\sum_{R_x} \frac{4\pi}{3}N(R_x) \sum_{\phi_p (R_x)} f_b((R_{s_p}+\Delta R_{s_p})^3-R_{s_p}^3) \sum_{\phi_q (R_x)} \phi_p (\phi_q -\phi_p) 	(1-f_i) \nonumber \\
			&\qquad\qquad [F(r,R_{s_p},R_{s_p}+\Delta R_{s_p},R_{s_q}) - F(r,R_{s_p},R_{s_p}+\Delta R_{s_p},R_{s_q}+\Delta R_{s_q})]\nonumber \\
			&\qquad - \sum_{R_x} \frac{4\pi}{3} N(R_x)\sum_{\phi_p (R_x)} f_b((R_{s_p}+\Delta R_{s_p})^3-R_{s_p}^3) \sum_{R'_x}\frac{4\pi}{3}N(R'_x) \sum_{\phi_q (R'_x)} f_b((R'_{s_q}+\Delta R'_{s_q})^3-{R'_{s_q}}^3) \nonumber \\
			&\qquad\qquad  \phi_p (\phi_q -\phi_p)\frac{1}{2}[F(r,R_{s_p},R_{s_p}+\Delta R_{s_p},R_h) + F(r,R'_{s_q},R'_{s_q}+\Delta R'_{s_q},R'_h)] \Bigg)
			 \label{eq:psi12fin}
	\end{align}
Where,
	\begin{align}
		\langle \phi_h \rangle &= \sum_{R_x} \frac{4\pi}{3} N(R_x) \sum_{\phi (R_x)} \!\! \phi  f_b ((R_s+\Delta R_s)^3-R^3_s) \nonumber \\
		\langle \phi_h^2 \rangle &= \sum_{R_x} \frac{4\pi}{3} N(R_x) \sum_{\phi (R_x)} \!\! \phi^2  f_b ((R_s+\Delta R_s)^3-R^3_s)
	\end{align}

We also calculate the  correlation at the same point using $\langle \phi_1 \phi_1\rangle = \langle \phi^2\rangle$,
	\begin{align}
		\mu_0 &= (1+\xi_0)(\phi_b^2 f_b + \langle \phi_h^2 \rangle) - (f_b\phi_b + \langle \phi_h \rangle)^2 
	\end{align}
where, $\xi_0 = \xi(r=0)$.
	
Eq~\ref{eq:psi12fin} matches with the expression derived in RS18 most of the cases. However, this expression is more robust as it gives expected results in all simplifying cases and boundary conditions.


\end{document}